\documentclass{aa}
\usepackage{natbib}
\bibpunct{(}{)}{;}{a}{}{,}
\usepackage{amsfonts}
\usepackage{psfrag}
\usepackage{graphicx}

\newcommand{\Msolar}{\mbox{\,$\rm M_{\odot}$}}        % solar mass
\begin{document}
\title{3D kinematics of white dwarfs from the SPY project II.%
\thanks{Based on observations obtained at the Paranal Observatory of the 
European Southern Observatory for programs 165.H-0588 and 167.D-0407}
\thanks{Tables 8 and 9 are only available in electronic form
at the CDS via anonymous ftp to cdsarc.u-strasbg.fr (130.79.128.5)
or via http://cdsweb.u-strasbg.fr/cgi-bin/qcat?J/A+A/}}

\author{E.-M. Pauli \inst{1}
\and R.Napiwotzki \inst{1,2,3}
\and U. Heber \inst{1}
\and M. Altmann \inst{1,4}
\and M. Odenkirchen \inst{5}
}
\institute{Dr.~Remeis-Sternwarte, Astronom.\ Institut, Universit\"at
        Erlangen-N\"urnberg, Sternwartstr.~7, 96049 Bamberg, Germany
\and
Department of Physics \& Astronomy, University of Leicester,
University Road, Leicester LE1 7RH, UK
\and
Centre for Astrophysics Research, University of Hertfordshire,
College Lane, Hatfield AL10 9AB, UK
\and
Departamento de Astronomia, Universidad de Chile, Camino Del
Observatorio 1515, Las Condes, Chile
\and
Max-Planck-Institut f\"ur Astronomie, K\"onigstuhl 17, 69117 Heidelberg,
 Germany
%\and
%Space Telescope European Coordinating Facility, ESO, Karl-Schwarzschild-Str.~2, 85748 Garching,
%        Germany
}
%\date{Received,  accepted }
\date{}
\offprints{U. Heber (heber@sternwarte.uni-erlangen.de)}

%%%%%%%%%%%%%%%%%%%%%%%%%%%%%%%%%%%%%%%%%%%%%%%%%%%%%%%%%%%%%%%%%%%%%%%
\abstract{
We present the kinematics of a sample of $398$ DA white dwarfs from the SPY 
project (ESO SN Ia Progenitor surveY) and discuss kinematic criteria for  
distinguishing of thin-disk, thick-disk, and halo populations. This is the 
largest homogeneous sample of white dwarfs for which 3D space motions have 
been determined. Since the percentage of old stars among white dwarfs is 
higher than among main-sequence stars, they are presumably valuable tools
in studies of old populations, such as the halo and the thick disk. Studies of 
white-dwarf kinematics can help to determine the fraction of the total mass of 
our Galaxy contained in the form of thick-disk and halo white dwarfs, an issue 
which is still under discussion. Radial velocities and spectroscopic distances 
obtained by the SPY project were combined with our measurements of proper 
motions to derive 3D space motions. Galactic orbits and further kinematic 
parameters were computed. We calculated individual errors of kinematic 
parameters by means of a Monte Carlo error propagation code. Our 
kinematic criteria for 
assigning population membership were deduced from a sample of F and G stars 
taken from the literature, for which chemical criteria can be used to 
distinguish between a thin-disk, a thick-disk and a halo star. 
Our kinematic population 
classification scheme is based on the position in the $U$\/-$V$-velocity 
diagram, the position in the $J_{\mathrm{z}}$-eccentricity diagram, and the 
Galactic orbit. We combined this with age information and found seven halo and
$23$ thick-disk white dwarfs in this brightness limited sample.
Another four rather cool white
dwarfs probably also belong to the thick disk. Correspondingly 2\% of the
white
dwarfs belong to the halo and 7\% to the thick disk. 
The mass contribution of the 
thick-disk white dwarfs 
is found to be substantial, but is insufficient to account for the
missing dark matter.
 
}

\maketitle
\section{Introduction: population membership of white dwarfs\label{intro}}
%%%%%%%%%%%%%%%%%%%%%%%%%%%%%%%%%%%%%%%%%%%%%%%%%%%%%%%%%%%%%%%%%%%%%%%%%%%

White dwarfs are the evolutionary end-products of most stars.
Since they are faint objects, only the nearby objects have been detected 
so far;
however, a large number of white dwarfs should be present in the Galaxy.
Determining the contribution of white dwarfs to the
total mass of the Galaxy
could help to solve one of the fundamental questions in
modern astronomy: what is the nature of dark matter?
The fact that the rotation curves of many galaxies are not Keplerian
\citep{rubin78}
invokes the existence of additional dark matter distributed in
a near-spherical structure, the so-called heavy-halo \citep{ostriker73}.
It is estimated that for the Milky Way only 10\% of the total
mass are present in the form of stars, gas, and dust in the Galactic disk
and halo \citep{alcock00}.
Dark matter candidates for the remaining 90\% include exotic particles,
cold molecular gas, and compact objects like black holes,
white dwarfs, and brown dwarfs.
The role of white dwarfs in the dark matter problem is still uncertain.
An open issue is the fraction of white dwarfs in the
thick-disk and halo populations, as well as their
fraction of the total mass of the Galaxy.
In this context, kinematic studies have proved a useful
tool in deciding on population membership of white dwarfs.

\citet{oppenheimer01} have claimed to have detected cool halo
white dwarfs as the main source for Galactic dark matter.
Their results have been discussed controversially by many groups:
e.g. \citet{reid01}, \citet{hansen01}, \citet{torres02}, and \citet{reyle01}, 
who criticise 
the input parameters used in the \citet{oppenheimer01} analysis.
The main problem lies in the lack of radial velocity information and 
poorly determined distances.
\citet{oppenheimer01} are criticised for putting radial velocities to zero, 
as well as for their distance estimates.

\citet{salim04} have overcome some of these problems by
measuring radial velocities for a subsample of those $13$ white dwarfs from
the \citet{oppenheimer01} sample that show an $H_{\rm \alpha}$ line.
They also present new CCD photometry for half of the sample,
in order to obtain new distance estimates.
Their new results confirm those of \citet{oppenheimer01}.
But it should be kept in mind that these results are based on small number
statistics. On the other hand, the velocity cut made 
by \citet{oppenheimer01}
was shown to be inappropriate for a proper motion limited survey 
\citep{reyle01,flynn03,graff01}. Moreover, \citet{bergeron03} 
found the white dwarfs of
\citet{oppenheimer01} to be warmer than 5000~K and therefore to most
likely be too young to belong to the halo.

The common problem of the investigations discussed above is the lack
of radial velocity measurements. Especially deviating conclusions derived
from the white dwarfs of the \citet{oppenheimer01} sample
demonstrate that different assumptions about the values of $v_{\rm rad}$
can produce different fractions of halo and thick-disk stars
and thus can affect the determination
of the white dwarf halo density.
Therefore a sample of white dwarfs with known radial velocity measurements
 is needed in
order to obtain the full 3D kinematic information.  

Examples of kinematics studies based on complete 3D space motions are
the samples from \citet{silvestri01,silvestri02}. In our previous study
\citep[][hereafter referred to as Paper~I]{pauli03},
we presented the first homogeneous
sample of white dwarfs for which complete 3D space motions were
determined thanks to precise radial velocities and spectroscopic
distances from high resolution 
%UH
spectra taken with UVES at the UT2 telescope of the ESO VLT.

Since radial velocities of white dwarfs are difficult to measure,
\citet{silvestri01,silvestri02} have
obtained radial velocities from the spectra of
the main-sequence companions of white dwarfs in common 
proper motion pairs. We 
derived white dwarf radial velocities from high resolution spectra directly 
and analysed a sample of $107$ single DA white dwarfs in Paper I. There we 
demonstrated 
how a combination of several kinematic classification
criteria allows efficient distinction of the different stellar populations.
In contrast to previous studies, we not only considered the classical velocity
components $U$, $V$, and $W$ of each white dwarf, but also calculated its orbit
in the
Galaxy.

This allowed us to define new sophisticated criteria for
classifying thin-disk, thick-disk, and halo populations by 
considering Galactic 
orbits and kinematic parameters.
Another important question is how errors of the input parameters affect
errors of the kinematic output parameters. An error propagation code
using a Monte Carlo simulation was developed which allowed us to check
the statistical significance of our results.
Four halo, and seven thick-disk white dwarfs were found.
Our fraction of halo white dwarfs is much smaller than the one of
\citet{oppenheimer01}, 
indicating that halo white
dwarfs are not a major component of the dark matter in the Galaxy.
%UH

In a next step, we enlarged the sample of white dwarfs analysed.
We present here a sample of $398$ DA white dwarfs from the
ESO {\bf S}N Ia
{\bf P}rogenitor surve{\bf Y} \citep[SPY,][]{napiwotzki01, napiwotzki03}.
The SPY sample allowed us to overcome several limitations of
previous investigations.
When investigating DA white dwarfs, radial velocities could be measured from
the shifts of the Balmer lines.
Due to high resolution UVES VLT spectra,
we could benefit from measurements of radial velocities of 
unprecedented precision
(typical errors of only $2\,{\rm km\,s^{-1}}$) and of spectroscopic distances
(relative errors of only 10\%). 
Effective temperatures and gravities are from \citet{koester01} 
and from preliminary results of Koester (priv. comm.). 
The final results will be reported in a forthcoming paper (Voss
et al., in preparation).

We supplemented these data with the best proper motion measurements available.
Therefore we possess a very homogeneous set of radial and tangential velocity
information with individual errors for each star.
We refined our population classification scheme, thanks to
a larger calibration sample, and applied it to $398$ SPY white dwarfs.

Our publication is structured as follows:
Sect.~\ref{data} deals with the input data.
In Sect.~\ref{orbit} our kinematic analysis method is described and applied to
the calibration sample and to the sample of SPY white dwarfs.
Our results appear in Sect.~\ref{popuclasswd} and are discussed in
Sects.~\ref{chap_ages} and \ref{dis}. We finish with conclusions in 
Sect.~\ref{con}.
%%%%%%%%%%%%%%%%%%%%%%%%%%%%%%%%%%%%%%%%%%%%%%%%%%%%%%%%%%%%%%%%%%%%%%%%%%%
%%%%%%%%%%%%%%%%%%%%%%%%%%%%%%%%%%%%%%%%%%%%%%%%%%%%%%%%%%%%%%%%%%%%%%%%%%%

\section{Kinematical data\label{data}}
\subsection{Sample \label{obs}}
Our sample consists of $398$ DA white dwarfs from the SPY project. 
We have to stress that our stars are drawn from a brightness limited 
selection of known white dwarfs 
\citep[see][for details]{napiwotzki01},
and are not selected from proper motion surveys.
Radial velocities, corrected for gravitational redshift were taken from
Napiwotzki et al. (in prep.) and spectroscopic distances from 
\citet{koester01} and Voss et al. (in  prep.).
Radial velocities were measured using the cross-correlation technique
described in \citet{napiwotzki01} but are still somewhat preliminary
(final results to be published in a forthcoming paper). Uncertainties in
the radial velocities are mostly much smaller than those of the tangential
velocities computed from the proper motions (see also Paper I). 

The aim of the SPY project is to detect radial velocity (RV) variable
binary white dwarfs. Two spectra at different epochs were taken
and checked for RV variations.
Since orbital motions distort the measurement of space motions,
RV variable stars were discarded from our sample.
It should be noted that, while  more than
one spectrum is available for most SPY white dwarfs, there are $52$ stars 
where only one spectrum exists.
Of the $107$ stars analysed in Paper~I, only $104$ are also
present in this larger $398$ sample. Three stars have turned out
(according to information from additional spectra) to be in binary systems
 so were excluded from the sample.
Given a binarity fraction of about $5\%$ \citep{napiwotzki05},
we can estimate that there may be only two or three still undetected
spectroscopic binaries in the kinematically analysed sample,
a negligible number.
Nevertheless, in Table~8 we mark those stars with an asterisk where the radial 
velocity
is based on only one spectrum.
%%%%%%%%%%%%%%%%%%%%%%%%%%%%%%%%%%%%%%%%%%%%%%%%%%%%%%%%%%%%%%%%%%%%%%%
\subsection{Proper motions}
Sources for proper motions are
the USNO--B catalogue \citep{monet03},
the SuperCOSMOS Sky Survey \citep{hambly01_3,hambly01_2,hambly01_1},
the UCAC2 catalogue \citep{zacharias00_1},
the Yale Southern Proper Motion catalogue \citep{girard03},
the revised NLTT catalogue \citep{salim03,gould03_1},
and the revised LHS catalogue \citep{bakos02}.
Additional proper motions (for $202$ stars) were measured using the
Bonner Astrometry Software \citep{geffert97}, the procedure is described
in Paper~I. 

For most programme stars, there is more than one astrometric measurement.
The question is how to combine the proper motions and their errors
from the different sources.
%%RN
If all the different measurements were completely 
independent of each other and if
the errors followed a Gaussian distribution, average proper motions
$\left<\mu\right>$ and their errors $\left<\sigma_{\rm \mu}\right>$
would have to be weighted by the inverse variances, as:
%%%%%%%%%%%%%%%%%%%%%%%%%%%%%%%%%%%%%%%
\begin{equation}
\left< \mu\right> =
{\left(\sum_{i=1}^{n}{\mu_{\rm i}/{\sigma^{2}_{\rm \mu i}}}\right)}
/
{\left(\sum_{i=1}^{n}{1/{\sigma^{2}_{\rm \mu i}}}\right)}\ .
\label{eqmu}
\end{equation}
%%%%%%%%%%%%%%%%%%%%%%%%%%%%%%%%%%%%%
\begin{equation}
\left< \sigma_{\rm \mu}\right> =
{\left(\sum_{i=1}^{n}{1/{\sigma^{2}_{\rm \mu i}}}\right)}^{-1/2}\ .
\label{eqsigmu}
\end{equation}
%%%%%%%%%%%%%%%%%%%%%%%%%%%%%%%%%%%%
We know, however, that not all measurements are independent of each
other, since some of the catalogues share the same plate material.
Furthermore, though we checked as far as possible whether the star
found in the catalogue by the
automatic search procedure using its coordinates
is indeed the white dwarf, in some cases misidentifications have occurred.
Comparing the proper motions of one star in different
sources permits false detections to be eliminated.

To do this, we calculated the combined average and error, plus the
quadratic deviation $\Delta^{2}_{\rm \mu i}$
of an individual measurement from this average:
%%%%%%%%%%%%%%%%%%%%%%%%%%
\begin{equation}
 \Delta^{2}_{\rm \mu i} =
\left( \mu_{\rm i}-\left< \mu\right> \right) ^{2}\ .
\end{equation}
%%%%%%%%%%%%%%%%%%%%%%%%%%%
If each $\Delta^{2}_{\rm \mu i}$ is divided by the corresponding
$\sigma_{\rm \mu i}^{2}$,
the sum over all $i$ is taken and divided by the number of measurements $n$.
We
get a quantity $\Delta_{\rm check}$ that allows to check if the individual
measurements are
consistent with each other and, if not, to eliminate the measurement
which differs from the others:
%%%%%%%%%%%%%%%%%%%%%%%%%%%%
\begin{equation}
\Delta_{\rm check}=
{1\over n}
\left(\sum_{i=1}^{n} \left( \Delta^{2}_{\rm \mu i}/\sigma^{2}_{\mu i}
\right) \right)\ .
\end{equation}
%%%%%%%%%%%%%%%%%%%%%%%%%%%
If $\Delta_{\rm check}>1$, we checked the different catalogue values
manually in order to decide which values to choose and which to eliminate.
Having thus eliminated false detections the next step was to calculate
the quantity $\left< \Delta_{\rm \mu}\right>$:
%%%%%%%%%%%%%%%%%%%%%%%%%%
\begin{equation}
\left< \Delta_{\rm \mu}\right> =
\sqrt
{{1\over n} \left(\sum_{i=1}^{n} \Delta^{2}_{\rm \mu i}\right)}\ .
\end{equation}
%%%%%%%%%%%%%%%%%%%%%%%%%
We adopted the weighted mean $\left< \mu\right>$ from Eq.~(\ref{eqmu}) 
and the maximum of $\left< \sigma_{\rm \mu}\right>$
and $\left< \Delta_{\rm \mu}\right>$ as
the corresponding error.
This enabled us to obtain a realistic error estimate,
which typically lies between $5\,{\rm mas~{yr}^{-1}}$ and
$10\,{\rm mas~{yr}^{-1}}$.
%%%%%%%%%%%%%%%%%%%%%%%%%%%%%%%%%%%%%%%%%%%%%%%%%%%%%%%%%%%%%%%%%%%%%%%%%%%%

The input parameters radial velocities, spectroscopic distances,
and
proper motion components together with their errors, are listed
for all white dwarfs in Table~8.
%%%%%%%%%%%%%%%%%%%%%%%%%%%%%%%%%%%%%%%%%%%%%%%%%%%%%%%%%%%%%%%%%%%%%%%%%%%%

%%%%%%%%%%%%%%%%%%%%%%%%%%%%%%%%%%%%%%%%%%%%%%%%%%%%%%%%%%%%%%%%%%%%%%%%%%
\section{Revised population classification scheme\label{orbit}}
In Paper~I we presented
a new sophisticated
population classification scheme based on the $U$\/-$V$-velocity diagram,
the $J_Z$-eccentricity-diagram, and the Galactic orbit.
For the computation of orbits and kinematic parameters, we used the code
by \citet{odenkirchen92} based on a Galactic potential by \citet{allen91}.
The classification scheme was based on a calibration sample of
main-sequence stars.
In the meantime, new spectroscopic analyses have become available which 
allowed us to enlarge the calibration sample and to refine our
classification criteria.
%%%%%%%%%%%%%%%%%%%%%%%%%%%%%%%%%%%%%%%%%%%%%%%%%%%%%%%%%%%%%%%%%%%%%%%%%%%

\subsection{The calibration sample\label{cal}}
Unlike for main-sequence stars, the population membership of white dwarfs
cannot be determined from spectroscopically measured metalicities.
Therefore we have to rely on kinematic criteria.
Those criteria have to be calibrated using a suitable calibration sample of
main-sequence stars.
In our case this sample consists of $291$ F and G main-sequence stars from
\citet{edvardsson93}, \citet{fuhrmann98},
Fuhrmann (2000\footnote{\tt http://www.xray.mpe.mpg.de/fuhrmann/.},
2004). It is important to note that the stars were 
selected from flux limited samples and not from proper motion surveys.
Thanks to the work of \citet{fuhrmann04},
the number of calibration sample stars has been doubled, which makes it 
worthwhile revisiting the classification criteria outlined in Paper I.

\begin{figure}
  \resizebox{\hsize}{!}{
\begin{psfrags}
\psfrag{[Fe/H]}{${\rm [Fe/H]}$}
\psfrag{[Mg/Fe]}{${\rm [Mg/Fe]}$}
\psfrag{thin disk}{thin disk}
\psfrag{thick disk}{thick disk}
\psfrag{halo}{halo}
\psfrag{trans}{trans}
\psfrag{metweak}{metweak}
    \includegraphics[width=17cm]{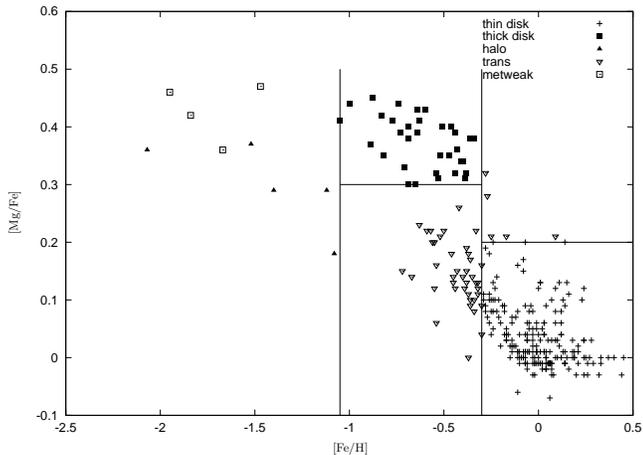}
\end{psfrags}}
  \caption{${\rm [Mg/Fe] vs. [Fe/H]}$ abundance diagram for the calibration 
sample (see text). 
}
\label{met}
\end{figure}

For both samples a detailed abundance analysis was carried out.
\citet{fuhrmann98} combined abundances, ages, and 3D kinematics for
population classification and found that the disk and halo populations
can be distinguished best
in the [Mg/Fe] versus [Fe/H] diagram. Halo and thick-disk stars can be
separated by means of their [Fe/H] abundances,
as they possess a higher [Mg/Fe] ratio than thin-disk stars
\citep[see also][]{bensby03}.
In Fig.~\ref{met} the
${\rm [Mg/Fe]}$ versus ${\rm [Fe/H]}$ abundances
for the $291$ main-sequence stars are shown.
These stars are divided into halo, thick disk, and
thin disk according to their position in the diagram.
The halo stars have $[{\rm Fe}/{\rm H}]<-1.05$,
the thick-disk stars
$-1.05\le[{\rm Fe}/{\rm H}]\le-0.3$ and
$[{\rm Mg}/{\rm Fe}]\ge 0.3$, and the thin-disk stars 
$[{\rm Fe}/{\rm H}]>-0.3$ and $[{\rm Mg}/{\rm  Fe}] \le 0.2$.
Stars in the overlapping area between the thin and the thick disk (open
triangles in Fig.~\ref{met}) were
neglected in order to ensure a clear distinction between the
two disk populations.

There are four stars left to the halo border, which according
to \citet{fuhrmann04} belong to the metal-weak thick disk
(MWTD, open boxes).
As their kinematics are indeed incompatible with halo
membership, we omitted them
from further analysis.
Also rejected was the star HD\,148816, which though in
the thick-disk region in the abundance diagram, clearly shows
halo kinematics (not shown in the diagram).

This demonstrates that a clear distinction between halo and
thick-disk stars by means of abundances is difficult,
but as will be shown later, halo and thick-disk stars
show very distinct kinematic
properties, so that they are unlikely to be confused.
%%%%%%%%%%%%%%%%%%%%%%%%%%%%%%%%%%%%%%%%%
%%%%%%%%%%%%%%%%%%%%%%%%%%%%%%%%%%%%%%%%%%%%%%%%%%%%%%%%%%%%%%%%%%%%%%%%%%%%%%%
\subsection{The $U$\/-$V$-velocity diagram\label{uv}}
\begin{figure*}
  \centering
\begin{psfrags}
\psfrag{V(kms^-1)}{$V/{\rm km~s^{-1}}$}
\psfrag{U(kms^-1)}{$U/{\rm km~s^{-1}}$}
\psfrag{thin disk}{thin disk}
\psfrag{thick disk}{thick disk}
\psfrag{halo}{halo}
\psfrag{3 sig thin}{$3\sigma_{\rm thin}$-limit}
\psfrag{3 sig thick}{$3\sigma-{\rm thick}$-limit}
    \includegraphics[width=17cm]{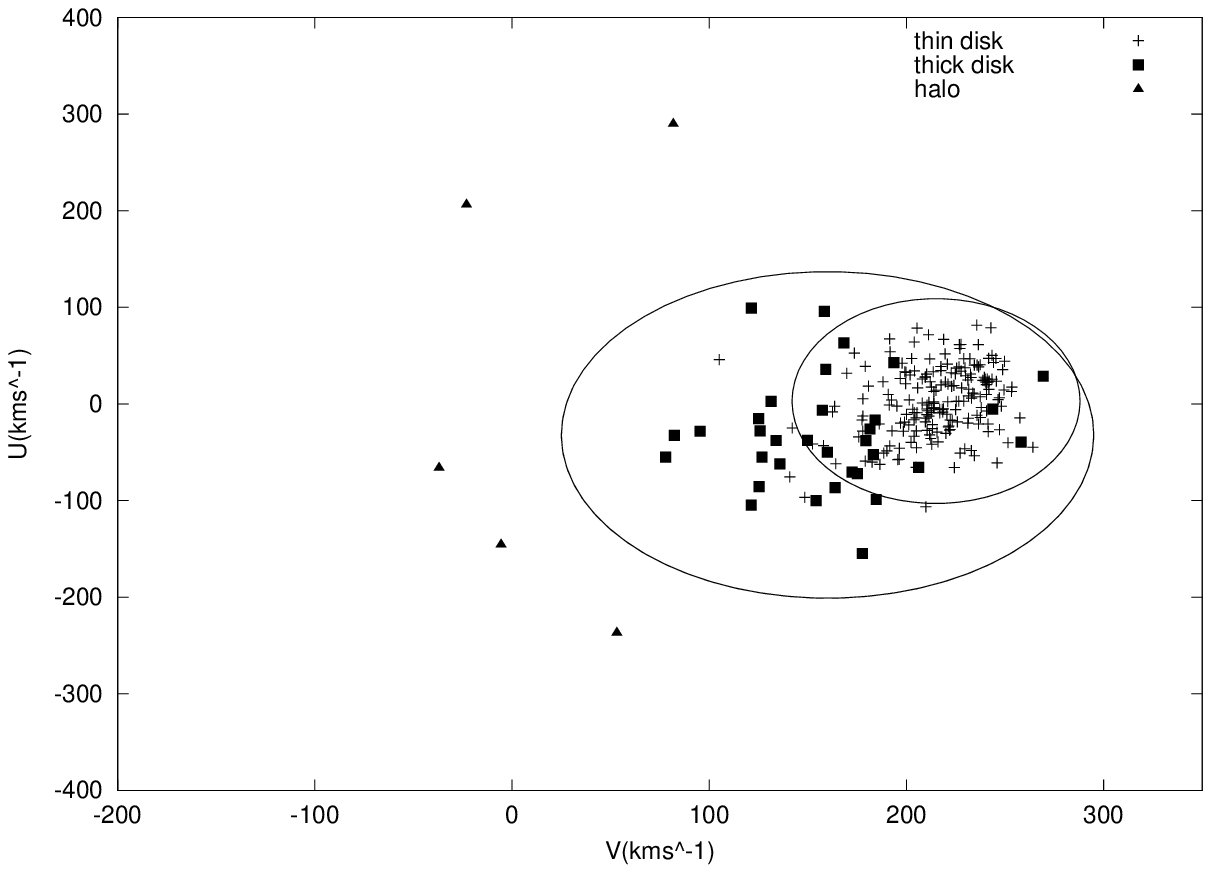}
\end{psfrags}
  \caption{$U$\/-$V$-velocity diagram for the calibration sample of 
  main-sequence stars with $3\sigma_{\rm thin}$-, $3\sigma_{\rm
thick}$-contours.}
\label{uvms}
%\end{figure*}
%%%%%%%%%%%%%%%%%%%%%%%%%%%%%%%%%%%%%%%%%%%%%%%%%%%%%%%%%%%%%%
%\begin{figure*}
  \centering
\begin{psfrags}
\psfrag{e}{$e$}
\psfrag{Jz(kpc km s^-1)}{$J_Z/{\rm kpc\,km~s^{-1}}$}
\psfrag{thin disk}{thin disk}
\psfrag{thick disk}{thick disk}
\psfrag{halo}{halo}
\psfrag{RegionA}{Region~A}
\psfrag{RegionB}{Region~B}
\psfrag{RegionC}{Region~C}
    \includegraphics[width=17cm]{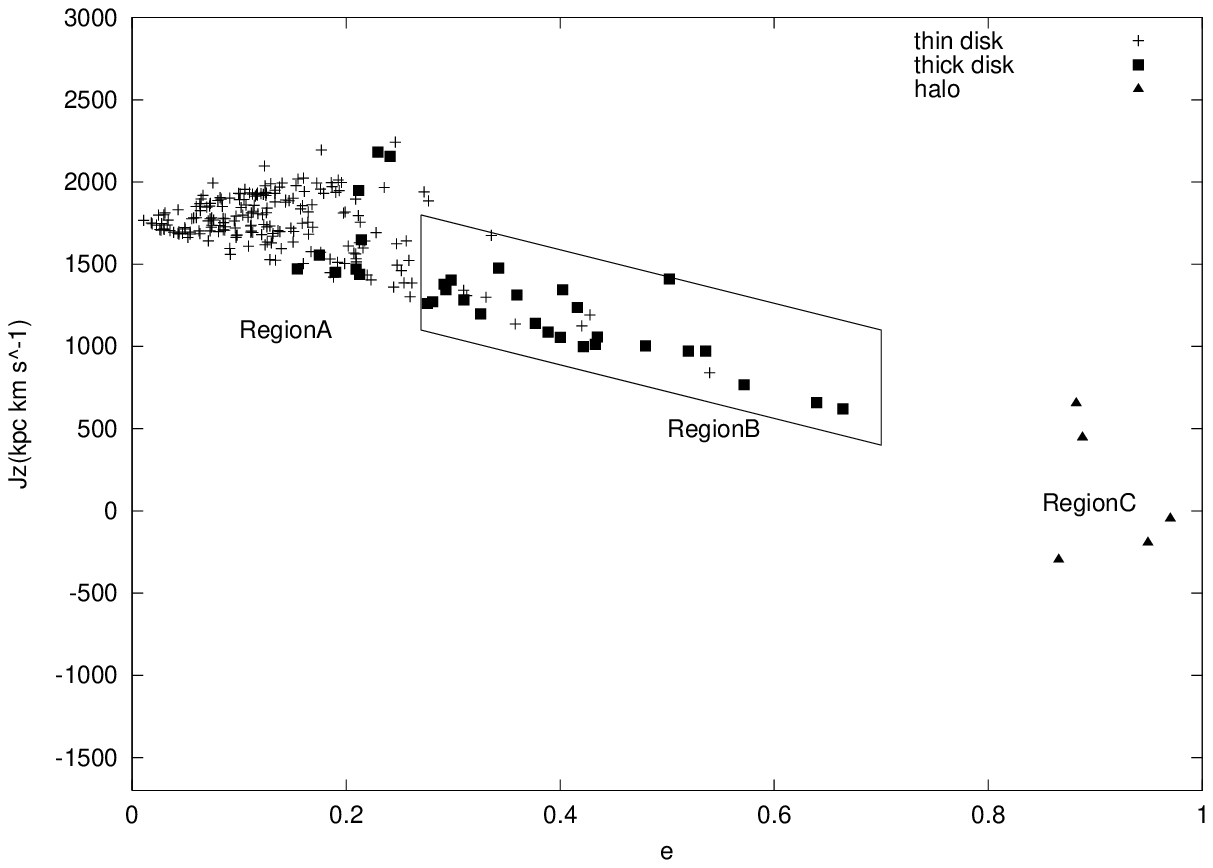}
\end{psfrags}
  \caption{$J_Z$-$e$-diagram for the calibration sample of main-sequence stars}
  \label{ecc1}
\end{figure*}

A classical tool for kinematic investigations is the
$U$\/-$V$-velocity diagram. In Fig.~\ref{uvms}, $U$ is plotted
versus $V$ for the main-sequence stars.
For the thin-disk and the thick-disk stars, the mean values and standard
deviations of the two velocity components were calculated.
The values for the thin disk are:
$\left<U_{\rm ms}\right>=3\,\rm{km~s^{-1}}$,
$\left<V_{\rm ms}\right>=215\,\rm{km~s^{-1}}$,
$\sigma_{U_{\rm ms}}=35\,\rm{km~s^{-1}}$, and
$\sigma_{V{\rm ms}}=24\,\rm{km~s^{-1}}$.
The corresponding values for the thick disk are:
$\left<U_{\rm ms}\right>=-32\,\rm{km~s^{-1}}$,
$\left<V_{\rm ms}\right>=160\,\rm{km~s^{-1}}$,
$\sigma_{U_{\rm ms}}=56\,\rm{km~s^{-1}}$, and
$\sigma_{V{\rm ms}}=45\,\rm{km~s^{-1}}$.
The negative value of $\left<U_{\rm ms}\right>$ 
is explained
in \citet{fuhrmann04} as an effect of the Galactic bar.
Indeed, nearly all thin-disk stars stay inside the
$3\sigma_{\rm thin}$-limit, and all halo stars lie outside
the $3\sigma_{\rm thick}$-limit, as can be seen from Fig.~\ref{uvms}.
In our previous paper, we used the $2\sigma-$limit of the thin and
thick-disk stars for finding thick-disk stars and
$\sqrt{U^2+(V-195)^2}\ge 150\,\rm{km\,s^{-1}}$ for finding halo stars.
We replaced these by the 
more stringent 
$3\sigma-$limits of the thin and thick-disk
stars to obtain a clear-cut separation. 
%%%%%%%%%%%%%%%%%%%%%%%%%%%%%%%%%%%%%%%%%%%%%%%%%%%%%%%%%%%%%
%%%%%%%%%%%%%%%%%%%%%%%%%%%%%%%%%%%%%%%%%%%%%%%%%%%%%%%%%%%%%%
\subsection{The $Jz$-$e$-diagram \label{jze}}
The $U$\/-$V$-plot is not the only source of information about population
membership.
Two important orbital parameters
are the $z$-component of the angular momentum $J_Z$ and the
eccentricity of the orbit $e$. Both are plotted against each other
for the main-sequence stars in Fig.~\ref{ecc1}.
The different populations can be distinguished well in this diagram.
The thin-disk stars cluster in a V-shaped area of
low eccentricity and $J_Z$ around $1\,800\,\rm{kpc \, km~s^{-1}}$,
%%RN
which we denote as region~A.

In general, the thick-disk stars possess higher
eccentricities $e>0.27$ and lower angular momenta.
They can be found in region~B.
There is also a clump of thick-disk
stars with lower eccentricity around $0.2$ and higher $J_Z$.
Region~B is defined such that it excludes as many thin-disk
stars as possible. The price that has to be paid for this is the
loss of some thick-disk stars. But this way there is a high probability
of identifying only those stars as thick-disk members that really
belong to the thick disk.
It should be noted that region~3 in our previous paper, which seemed
to be different from the thin-disk and the thick-disk regions A and B, has
proven to be just an extension of the thin-disk region to higher
eccentricities. Therefore it does not appear as an additional region in
this revised classification scheme.

The halo stars with very high eccentricity and smaller
$J_Z$ can be found in Region~C,  separated well from all other stars.
%%%%%%%%%%%%%%%%%%%%%%%%%%%%%%%%%%%%%%%%%%%%%%%%%%%%%%%%%%%%%%%%%%%%%%%%
\subsection{Galactic orbits\label{orb}}
The eccentricity was extracted from the Galactic orbit of the stars.
The classification can be confirmed by checking
the orbits themselves.
Typical orbits for thin-disk, thick-disk and halo main-sequence stars
can be found 
in Paper~I and will not be repeated here.
%%%%%%%%%%%%%%%%%%%%%%%%%%%%%%%%%%%%%%%%%%%%%%%%%%%%%%%%%%%%%%%%%%%%%%%
\subsection{Population classification scheme \label{class}}

Our classification scheme (developed in Paper~I) combines three
different classification criteria:
i) the position in $U$\/-$V$ diagram,
ii) the position in $J_Z$-$e$ diagram, and iii) the Galactic orbit.

We repeat some details here of the population classification scheme presented 
in Paper~I and then describe the new refinements and changes.
We classified white dwarfs as halo members if they had a value of
$\sqrt{U^2+(V-195)^2}\ge 150\,\rm{km\,s^{-1}}$ and 
lay in region~4 in the $J_Z$-$e$-diagram (see Paper~I).

To detect thick-disk white dwarfs, first all stars either
situated outside the $2\sigma$-limit
in the $U$\/-$V$-diagram or in region~2 or 3 in the $J_Z$-$e$-diagram
were selected as thick-disk candidates.
In a second step, each candidate was assigned a classification value
$c$. $c$ was defined as the sum of the individual values $c_{\rm UV}$,
$c_{\rm J_{Z}e}$ and $c_{\rm orb}$ corresponding to the three
different criteria: position in $U$\/-$V$-diagram,
position in $J_Z$-$e$-diagram, and Galactic orbit.

We assigned $c_{\rm UV}=+1$ to a star outside the $2\sigma$-limit
in the $U$\/-$V$-diagram, whereas one inside the $2\sigma$-limit
got $c_{\rm UV}=-1$.
The different regions in the $J_Z$-$e$-diagram are characterised by
$c_{\rm J_{Z}e}=-1$ for region~1, $0$ for region~3, and $+1$ for region~2.
The third classification value $c_{\rm orb}$ described the orbits:
$c=-1$ for orbits of thin-disk type and $c=+1$ for orbits of
thick-disk type.
Then the sum $c=c_{\rm UV}+c_{\rm J_{Z}e}+c_{\rm orb}$ was computed.
Stars with $c=+3$  or $c=+2$ were considered as bona fide
thick-disk members, and those with $c=+1$ as probable thick-disk members.
If $c \le 0$, the star was classified as belonging to the thin disk.

The new classification scheme is more concise due to the elimination of
region~3. 
As described in Sect.~\ref{uv}, we also sharpened the selection criterion 
for the  $U$\/-$V$ plane by replacing the $2\sigma$ by a $3\sigma$ limit.
A star is classified as a halo candidate if it lies either
outside the $3\sigma_{\rm thick}$-limit in the $U$\/-$V$ diagram
or in region~C in the $J_Z$-$e$ diagram.
Then classification values $c_{\rm UV}$, $c_{\rm J_{Z}e}$, and $c_{\rm orb}$
are assigned to all halo candidates
which take the value of $+1$ if the criterion favors a halo
membership and $-1$ if not.
More precisely:
$c_{\rm UV}=+1$ if the star lies outside the $3\sigma_{\rm thick}$-limit,
$c_{\rm J_{Z}e}=+1$ if the star lies in region~C,
and $c_{\rm orb}=+1$ if the star has a halo orbit.
Then the sum $c=c_{\rm UV}+c_{\rm J_{Z}e}+c_{\rm orb}$ is calculated.
All of the halo candidates with $c \ge +1$
are classified as halo members, the rest as thick-disk members.

All the remaining stars (not found to belong to the halo), 
either outside the $3\sigma_{\rm thin}$-limit in the $U$\/-$V$ diagram
or in region~B in the $J_Z$-$e$ diagram, are classified
as thick-disk candidates.
Then the analogous procedure to the halo classification is applied:
$c_{\rm UV}=+1$ if the star lies outside the $3\sigma_{\rm thin}$-limit,
$c_{\rm J_{Z}e}=+1$ if the star lies in region~B,
and $c_{\rm orb}=+1$ if the star has a thick-disk orbit.
In contrast to Paper~I due to the elimination of region~3,
there is no longer a value 0 to be assigned
to $c_{\rm J_{Z}e}$; hence, we expect the number of thick-disk
candidates to decrease.
All of the thick-disk candidates with $c \ge +1$
are assigned to the thick-disk population, the rest to the thin-disk
population.
%%%%%%%%%%%%%%%%%%%%%%%%%%%%%%%%%%%%%%%%%%%%%%%%%%%%%%%%%%%%%%%%%%%%%
\subsection{Consistency check for the kinematical classification 
criteria \label{consist}}
In this section a consistency check of our classification
scheme is performed.
This is done by applying our kinematic classification criteria to
our calibration main-sequence sample.

Thirty-three main-sequence stars are known to belong to the thick disk because 
of their abundance patterns
(For reasons mentioned above we have excluded here the metal-weak thick-disk
stars), and
22 of them have a kinematical classification value $c \geq +1$ and
are classified as thick-disk stars.
Only one of them has $c=0$ and is thus misclassified as a thin-disk star.
This corresponds to a detection efficiency  of about $67\%$
for thick-disk members.
In addition to those 22 stars, six thin-disk main-sequence stars
with $c \geq +1$ are misclassified as thick-disk stars, so that
the total number of stars classified as thick disk is 28 
indicating a contamination with thin-disk stars of about $21\%$.

\subsection{Application to the white dwarf sample of Paper~I
\label{application}}

Furthermore, in order to be able to compare the results of Paper~I with
this paper, we applied the new classification scheme to the
$107$ white dwarfs analysed in Paper~I.
The fraction of halo stars is not changed by this new scheme.
Due to the elimination of region~3 in the $J_Z$-$e$ diagram, four stars
lose their thick-disk candidate status, and we end up with a
total number of eight thick-disk stars compared to twelve previously.
This reduces the local fraction of thick-disk white dwarfs from
$11\%$ to $7.5\%$, and
demonstrates the uncertainty of kinematic population classification.
Even higher errors are to be expected when the population separation
is based on a single criterion such as the position in $U$\/-$V$ diagram alone,
which is the case for most other kinematical studies of white dwarfs in the
literature.

%%%%%%%%%%%%%%%%%%%%%%%%%%%%%%%%%%%%%%%%%%%%%%%%%%%%%%%%%%%%%%%%%%%%%%%%%%%%%%
%%%%%%%%%%%%%%%%%%%%%%%%%%%%%%%%%%%%%%%%%%%%%%%%%%%%%%%%%%%%%
%%%%%%%%%%%%%%%%%%%%%%%%%%%%%%%%%%%%%%%%%%%%%%%%%%%%%%%%%%%%
\section{Kinematic population classification of the SPY white 
dwarfs\label{popuclasswd}}

We calculated orbits and kinematic parameters for all $398$ white
dwarfs (see Table~9
%\ref{kinpar2}
).
The errors of $e$, $J_Z$, $U$, $V$, $W$ were
computed with the Monte Carlo error propagation code
described in Paper~I.
They can be found in Table~9
%\ref{kinpar2} 
as well.
%%%%%%%%%%%%%%%%%%%%%%%%%%%%%%%%%%%%%%%
\begin{figure*}
  \centering
\begin{psfrags}
\psfrag{V(kms^-1)}{$V/{\rm km~s^{-1}}$}
\psfrag{U(kms^-1)}{$U/{\rm km~s^{-1}}$}
\psfrag{HE0201}{\footnotesize HE\,0201}
\psfrag{HS1527}{\footnotesize HS\,1527}
\psfrag{WD0252}{\footnotesize WD\,0252}
\psfrag{WD1448}{\footnotesize WD\,1448}
\psfrag{WD1524}{\footnotesize WD\,1524}
\psfrag{WD2351}{\footnotesize WD\,2351}
\psfrag{WD2359}{\footnotesize WD\,2359}
\psfrag{WD2029}{\footnotesize WD\,2029}
\psfrag{3 sig thin}{$3\sigma-{\rm thin}$-limit}
\psfrag{3 sig thick}{$3\sigma-{\rm thick}$-limit}
    \includegraphics[width=17cm]{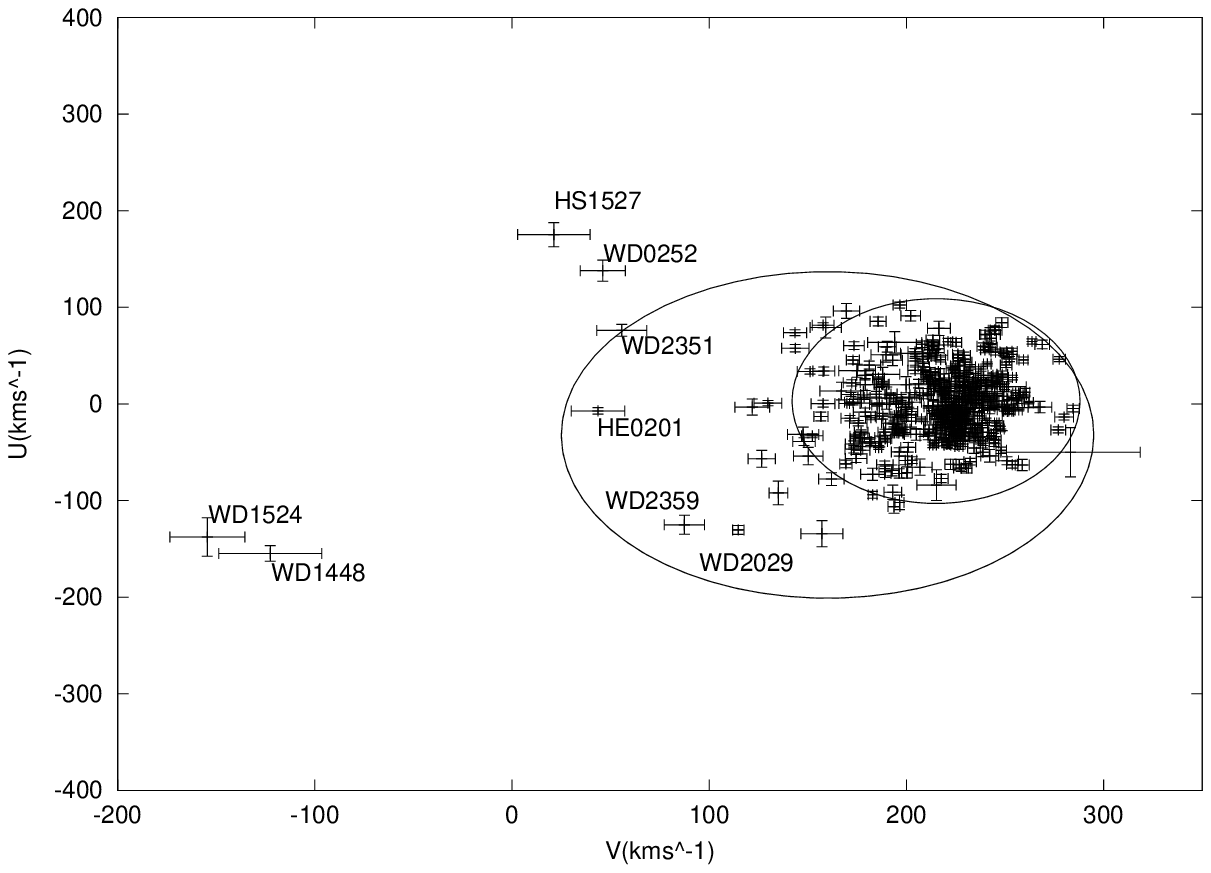}
\end{psfrags}
  \caption{$U$\/-$V$-velocity diagram for the white dwarfs with
$3\sigma-{\rm thin}$ and $3\sigma-{\rm thick}$ -- contours from 
Fig.~\ref{uvms}, Symbols with numbers are the white dwarfs 
mentioned in the text}
  \label{uvwd}
%\end{figure*}
%%%%%%%%%%%%%%%%%%%%%%%%%
%\begin{figure*}
  \centering
\begin{psfrags}
\psfrag{e}{$e$}
\psfrag{Jz(kpc km s^-1)}{$J_Z/{\rm kpc\,km~s^{-1}}$}
\psfrag{thin disk}{thin disk}
\psfrag{thick disk}{thick disk}
\psfrag{halo}{halo}
\psfrag{RegionA}{Region~A}
\psfrag{RegionB}{Region~B}
\psfrag{RegionC}{Region~C}
\psfrag{HE0201}{\footnotesize HE\,0201}
\psfrag{HS1527}{\footnotesize HS\,1527}
\psfrag{WD0252}{\footnotesize WD\,0252}
\psfrag{WD1448}{\footnotesize WD\,1448}
\psfrag{WD1524}{\footnotesize WD\,1524}
\psfrag{WD2351}{\footnotesize WD\,2351}
\psfrag{WD2359}{\footnotesize WD\,2359}
\psfrag{WD2029}{\footnotesize WD\,2029}
    \includegraphics[width=17cm]{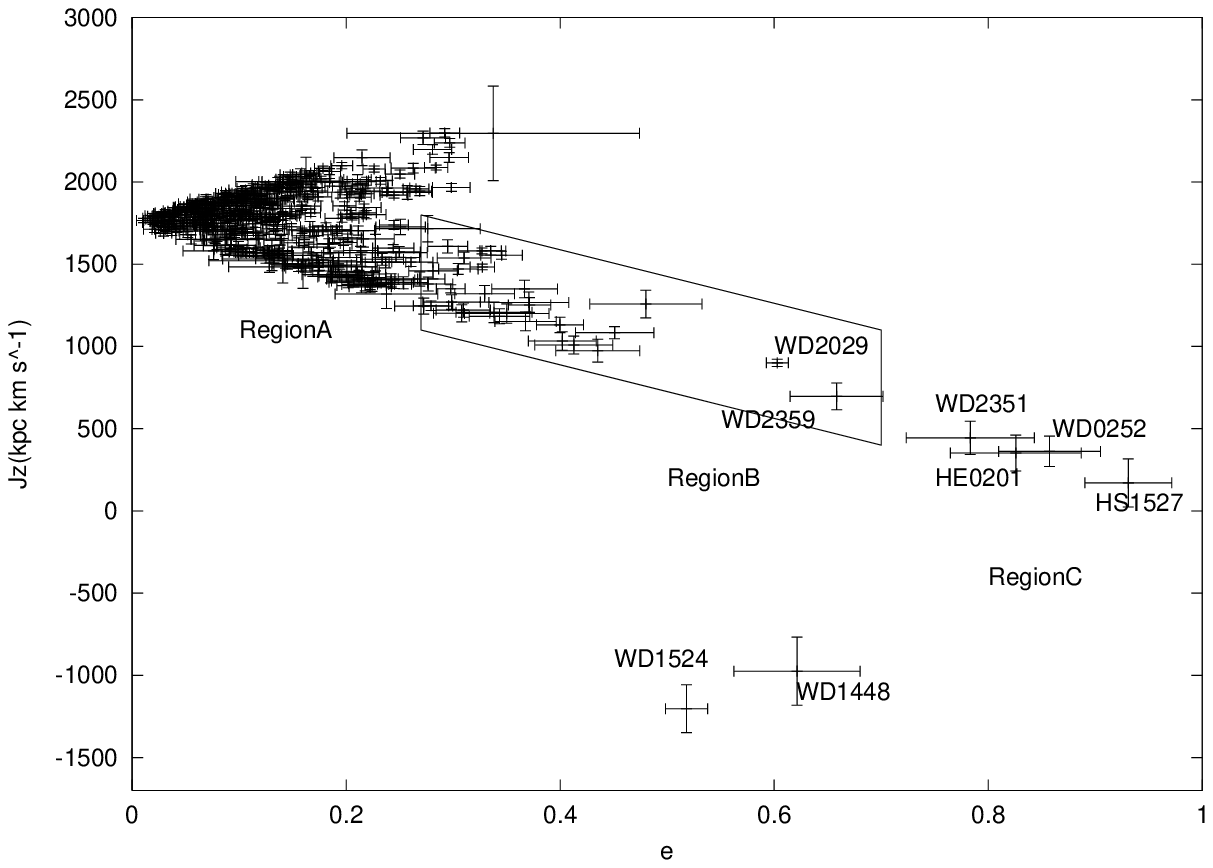}
\end{psfrags}
  \caption{$Jz$-$e$-diagram of the white dwarfs}
  \label{ecc2}
\end{figure*}
%%%%%%%%%%%%%%%%%%%%%%%%%%%%%%%%%%%%%%%%%%%%%%%%%%%%%%%%%%%%%%
%%%%%%%%%%%%%%%%%%%%%%%%%%%%%%%%%%%%%%%%%%%%%%%%%%%%%%%%%%%%%%
\subsection{The $U$\/-$V$-velocity diagram\label{uv_wd}}
In Fig.~\ref{uvwd}, the $U$\/-$V$-velocity diagram for the
white dwarfs is shown together with the $3\sigma$-limits
of the thin and thick-disk stars from the calibration sample.
The white dwarfs can be divided into two main groups that appear
to be separated from each other:
one group that is clustered mainly within the $3\sigma_{\rm thin}$-limit
with some stars just outside the $3\sigma_{\rm thin}$-border
and another second group with smaller $V$ that lies outside or just
inside the $3\sigma_{\rm thick}$-border.
All the white dwarfs belonging to the second group are marked with
the first letters of their names in Fig.~\ref{uvwd}.

The second group comprises five stars outside the
$3\sigma_{\rm thick}$-limit (which qualify as halo candidates according to
Sect.~\ref{class}) HS\,1527+0614, WD\,0252$-$350,
WD\,1448+077, WD\,1524$-$749, and WD\,2351$-$365.
 Exceptional are WD\,1448+077 and WD\,1524$-$749, which have
a negative value of $V$; i.e. they move on retrograde orbits.
This behaviour is incompatible with disk membership and
strongly suggests that they belong to the halo.

The other three white dwarfs of the second group are HE\,0201$-$0513,
WD\,2029+183 and WD\,2359$-$324. Situated inside the
$3\sigma_{\rm thick}$, they do not qualify as halo candidates but
we must check if they belong to the halo or to the thick disk by means of 
the $J_Z$-eccentricity diagram and the orbits .
%%%%%%%%%%%%%%%%%%%%%%%
\subsection{The $J_z$-$e$-diagram \label{jze_wd}}
We now move on to the $J_Z$-eccentricity diagram of the SPY
white dwarfs (Fig.~\ref{ecc2}).
Again, two groups of stars can be detected:
one first group starting in Region~A with a high-eccentricity tail in
Region~B, which represents the disk population, and a second
group in the right part of Region~B and in Region~C.
Contrary to the main-sequence stars there is a gap in Region~B
that is not populated at all by white dwarfs.
If this is real or just due to selection effects cannot be said at this point.

The second group contains all the stars discussed individually in the previous
section and labeled by name in Fig.~\ref{ecc2}.
HE\,0201$-$0513, since situated in Region~C, is
added to the list of halo candidates.
The two retrograde stars, WD\,1448+077 and WD\,1524$-$749,
can be distinguished easily by their negative value of $J_{\rm Z}$.
%%%%%%%%%%%%%%%%%%%%%%%%%%%%%%%%
\subsection{Galactic orbits}

\begin{figure}
  \resizebox{\hsize}{!}{
    \begin{psfrags}
      \psfrag{rho/kpc}{$\rho/{\rm kpc}$}
      \psfrag{Z/kpc}{$Z/{\rm kpc}$}
      \includegraphics{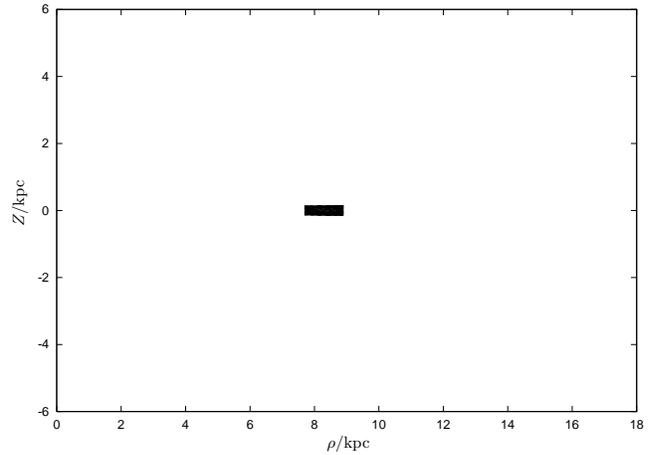}
    \end{psfrags}
    }
  \caption{WD\,0310$-$688: a white dwarf with a thin-disk orbit}
  \label{wd0310}
\end{figure}

\begin{figure}
  \resizebox{\hsize}{!}{
    \begin{psfrags}
      \psfrag{rho/kpc}{$\rho/{\rm kpc}$}
      \psfrag{Z/kpc}{$Z/{\rm kpc}$}
      \includegraphics{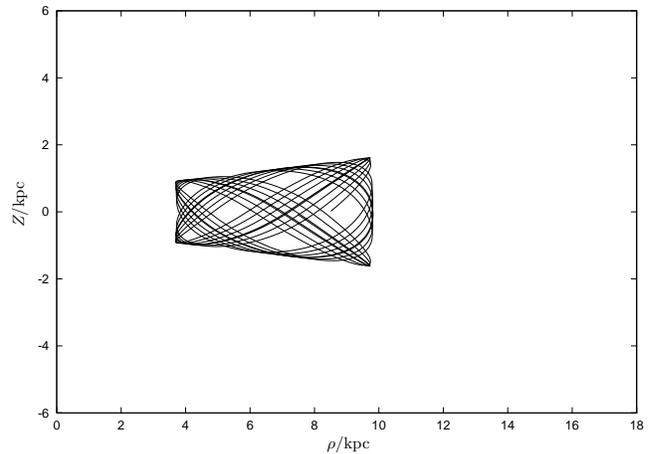}
    \end{psfrags}
    }
  \caption{WD\,1013$-$010: a white dwarf with a thick-disk orbit}
  \label{wd1013}
\end{figure}

\begin{figure}
  \resizebox{\hsize}{!}{
    \begin{psfrags}
      \psfrag{rho/kpc}{$\rho/{\rm kpc}$}
      \psfrag{Z/kpc}{$Z/{\rm kpc}$}
      \includegraphics{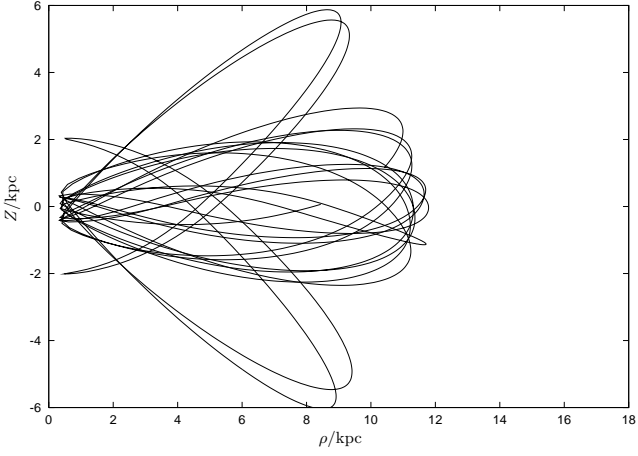}
   \end{psfrags}
    }
  \caption{HS\,1527+0614: a white dwarf with a (chaotic) halo orbit.} 
  \label{hs1527}
\end{figure}

Next we inspect the Galactic orbits of the SPY white dwarfs.
We display some meridional plots of white dwarfs with thin-disk, thick-disk, 
or halo like orbits, respectively, in Figs. \ref{wd0310} to \ref{hs1527}.

Most white dwarfs have thin-disk-like orbits, an example is
WD\,0310$-$688 (Figure~\ref{wd0310}).
Some orbits, like the one of WD\,1013$-$010 (Fig.~\ref{wd1013}),
show thick-disk characteristics.
The star WD\,2029+183 mentioned earlier has a thick-disk orbit.
Five stars (HS\,1527+0614,
HE\,0201$-$0513, WD\,0252$-$350, WD\,2351$-$365, and WD\,2359$-$324)
have chaotic halo orbits, as can be seen from 
Fig.~\ref{hs1527} in the case of 
HS\,1527+0614.
%%%%%%%%%%%%%%%%%%%%%%%%%%%%%%%%%%%%%%%%%%%%%%%%%%%%%%%%%%%%%

%%%%%%%%%%%%%%%%%%%%%%%%%%%%%%%%%%%%%%%%%%%%%%%%%%%%%%%%%%%%%%%%
%%%%%%%%%%%%%%%%%%%%%%%%%%%%%%%%%%%%%%%%%%%%%%%%%%%%%%%%%%%%%
%%%%%%%%%%%%%%%%%%%%%%%%%%%%%%%%%%%%%%%%%%%%%%%%%%%%%%%%%%%%%%%%%

\subsection{Classification}

We used the population classification scheme presented
in Sect.~\ref{class} to divide the SPY white dwarfs into
the three different populations. We start with the halo candidates, 
e.g. with all white dwarfs that are either
situated outside the $3\sigma$-limit of the thick disk in the
$U$\/-$V$-velocity diagram or that lie in Region~C in
the $J_Z$-eccentricity diagram.
Six white dwarfs fulfill these conditions:
all but one lie outside the $3\sigma_{\rm thick}$-limit, 
and all lie in Region~C.
Two white dwarfs, WD\,1448+077 and WD\,1524$-$749,
are on retrograde orbits characterised by a negative value
of $V$ and $J_{\rm Z}$.
When the classification values of the halo white
dwarf candidates are added, it is found that all of them have $c>1$ and
therefore belong to the halo population.
We have mentioned before that the star WD\,2359$-$324, though it
does not fulfill the criteria for a halo candidate, has an
orbit typical for a halo object.
As its error-bar places it near Region~C in the
$J_Z$-$e$ diagram, we therefore decided to classify it as a halo object.
This leaves us with seven halo white dwarfs.
Details can be found in Table~\ref{ha_class}.

We now move on to the remaining $32$ white dwarfs that lie either
outside the $3\sigma$-limit of the thin disk in the
$U$\/-$V$-velocity diagram or that lie in Region~B in
the $J_Z$-eccentricity diagram.
Twenty-seven of them have a classification value of $c>1$
and are classified as thick-disk members, the remaining
five are assigned a thin-disk membership (see Table~\ref{di_class}).
All the remaining white dwarfs are assumed to belong to the thin disk,
leaving us with seven halo, $27$ thick-disk, and
$364$ thin-disk out of the $398$ SPY white dwarfs.

%%%%%%%%%%%%%%%%%%%%%%%%%%%%%%%%%%%%%%%%%%%%%%%%%%%%%%%%%%%%%%
%table ha_class
%\input{2730tab1.tex}
\begin{table}
\caption[]
{Classification values for the halo candidates. Note that WD2359$-$324 is 
classified as a halo star despite having $c=-1$; see text \label{ha_class}}
\begin{tabular}{lrrrrl}
\\
\hline
star & $c_{\rm UV}$ & $c_{\rm J_Z-e}$ & $c_{\rm orb}$ & $c$ & classification\\ 
\hline
HE\,0201$-$0513 & \hspace*{.3cm} $-$1 & +1 & +1 & +1 & \hspace*{.45cm} 
halo \\
HS\,1527+0614 & \hspace*{.3cm} +1 & +1 & +1 & +3 & \hspace*{.45cm}     halo \\
WD\,0252$-$350 & \hspace*{.3cm} +1 & +1 & +1 & +3 & \hspace*{.45cm}     halo \\
WD\,1448+077 & \hspace*{.3cm} +1 & +1 & $-$1 & +1 & \hspace*{.45cm}     halo \\
WD\,1524$-$749 & \hspace*{.3cm} +1 & +1 & $-$1 & +1 & \hspace*{.45cm}   halo \\
WD\,2351$-$368 & \hspace*{.3cm} +1 & +1 & +1 & +3 & \hspace*{.45cm}     halo \\
WD\,2359$-$324 & \hspace*{.3cm} $-$1 & $-$1 & +1 & $-$1 & \hspace*{.45cm} 
halo \\
\hline
\end{tabular}
\end{table}

%table di_class
%\input{di_class.tex}
\begin{table}
\caption[]
{Classification values for the thick-disk candidates.  
\label{di_class}}
\begin{tabular}{lrrrrl}
\\
\hline
star & $c_{\rm UV}$ & $c_{\rm J_Z-e}$ & $c_{\rm orb}$ & $c$ & class.\\ 
\hline
HE\,0409$-$5154 & \hspace*{.3cm} +1 & +1 & $-$1 & +1 & \hspace*{.0cm} thick 
disk \\
HE\,0416$-$1034 & \hspace*{.3cm} +1 & $-$1 & +1 & +1 & \hspace*{.0cm} thick 
disk \\
HE\,0452$-$3444 & \hspace*{.3cm} +1 & +1 & +1 & +3 & \hspace*{.0cm} thick 
disk \\
HE\,0508$-$2343 & \hspace*{.3cm} +1 & +1 & $-$1 & +1 & \hspace*{.0cm} thick 
disk \\
HE\,1124+0144 & \hspace*{.3cm} +1 & +1 & +1 & +3 & \hspace*{.0cm} thick disk \\
HS\,0820+2503 & \hspace*{.3cm} +1 & +1 & +1 & +3 & \hspace*{.0cm} thick disk \\
HS\,1338+0807 & \hspace*{.3cm} $-$1 & +1 & +1 & +1 & \hspace*{.0cm} thick 
disk \\
HS\,1432+1441 & \hspace*{.3cm} $-$1 & +1 & +1 & +1 & \hspace*{.0cm} thick 
disk \\
WD\,0204$-$233 & \hspace*{.3cm} $-$1 & +1 & +1 & +1 & \hspace*{.0cm} thick 
disk \\
WD\,0255$-$705 & \hspace*{.3cm} +1 & +1 & +1 & +3 & \hspace*{.0cm} thick 
disk \\
WD\,0352+052 & \hspace*{.3cm} $-$1 & +1 & +1 & +1 & \hspace*{.0cm} thick 
disk \\
WD\,0548+000 & \hspace*{.3cm} $-$1 & +1 & $-$1 & $-$1 & \hspace*{.0cm} 
\it{thin disk} \\
WD\,0732$-$427 & \hspace*{.3cm} +1 & +1 & +1 & +3 & \hspace*{.0cm} thick 
disk \\
WD\,0956+045 & \hspace*{.3cm} $-$1 & +1 & $-$1 & $-$1 & \hspace*{.0cm} 
\it{thin disk} \\
WD\,1013$-$010 & \hspace*{.3cm} +1 & +1 & +1 & +3 & \hspace*{.0cm} thick 
disk \\
WD\,1152$-$287 & \hspace*{.3cm} $-$1 & +1 & +1 & +1 & \hspace*{.0cm} thick 
disk \\
WD\,1323$-$514 & \hspace*{.3cm} +1 & +1 & +1 & +1 & \hspace*{.0cm} thick 
disk \\
WD\,1327$-$083 & \hspace*{.3cm} +1 & +1 & $-$1 & +1 & \hspace*{.0cm} thick 
disk \\
WD\,1334$-$678 & \hspace*{.3cm} +1 & +1 & $-$1 & +1 & \hspace*{.0cm} thick 
disk \\
WD\,1410+168 & \hspace*{.3cm} +1 & +1 & +1 & +3 & \hspace*{.0cm} thick 
disk \\
WD\,1426$-$276 & \hspace*{.3cm} +1 & +1 & +1 & +3 & \hspace*{.0cm} thick 
disk \\
WD\,1507+021 & \hspace*{.3cm} $-$1 & +1 & +1 & +1 & \hspace*{.0cm} thick 
disk \\
WD\,1531+184 & \hspace*{.3cm} $-$1 & +1 & $-$1 & $-$1 & \hspace*{.0cm} 
\it{thin disk} \\
WD\,1614$-$128 & \hspace*{.3cm} +1 & +1 & $-$1 & +1 & \hspace*{.0cm} 
thick disk \\
WD\,1716+020 & \hspace*{.3cm} +1 & +1 & +1 & +3 & \hspace*{.0cm} thick 
disk \\
WD\,1834$-$781 & \hspace*{.3cm} +1 & +1 & +1 & +3 & \hspace*{.0cm} thick 
disk \\
WD\,1952$-$206 & \hspace*{.3cm} $-$1 & +1 & +1 & +1 & \hspace*{.0cm} thick 
disk \\
WD\,2029+183 & \hspace*{.3cm} +1 & +1 & +1 & +3 & \hspace*{.0cm} thick 
disk \\
WD\,2136+229 & \hspace*{.3cm} $-$1 & +1 & $-$1 & $-$1 & \hspace*{.0cm} 
\it{thin disk} \\
WD\,2253$-$081 & \hspace*{.3cm} $-$1 & +1 & $-$1 & $-$1 & \hspace*{.0cm} 
\it{thin disk} \\
WD\,2322$-$181 & \hspace*{.3cm} $-$1 & +1 & +1 & +1 & \hspace*{.0cm} thick 
disk \\
WD\,2350$-$083 & \hspace*{.3cm} $-$1 & +1 & +1 & +1 & \hspace*{.0cm} thick 
disk \\
\hline
\end{tabular}
\end{table}

%%%%%%%%%%%%%%%%%%%%

\section{Age estimates \label{chap_ages}}

The seven halo and $27$ thick-disk white dwarfs were assigned
to the respective populations by means of purely kinematic
criteria. Accordingly they must be old stars; therefore, we attempted to 
estimate their ages.
A check to see whether their physical parameters, mass and effective
temperature, are compatible with their belonging to an old population
must now be made. Masses $M$ for the white dwarfs were derived
from ${\rm log}~g$ and the mass-radius relation by \citet{wood95}.

The halo is older than $10\,{\rm Gyr}$.
\cite{bensby03} determined a mean age for the thick disk as
$11.2 \pm 4.3\,{\rm Gyr}$.
It is very probable that stars that are younger than $7\,{\rm Gyr}$
do not belong to the thick disk.
Thus the main-sequence life-time (plus about $20\%$ for time spent 
during the giant phases and the horizontal branch),
plus the time the white dwarf has cooled down
until it reaches its actual $T_{\rm eff}$, has to be greater than
the age of the youngest stars of the respective populations.
The main-sequence life-time $\tau_{\rm ms}$ depends on the mass of the
white dwarf progenitor and is approximately proportional to
$\tau_{\rm ms} \propto M^{-2.5}$ \citep{kippenhahn94}.
The main-sequence life-time is $\tau_{\rm ms}=10\,{\rm Gyr}$ for the Sun,
$7.9\,{\rm Gyr}$ for a $1.1\,\Msolar$ mass star,
$4.3\,{\rm Gyr}$ for a $1.4\,\Msolar$ mass star, and
$1.8\,{\rm Gyr}$ for a $2\,\Msolar$ mass star.
Adding the $20\%$ horizontal branch plus giant phase lifetime, the total
pre-white dwarf lifetimes would be
$12\,{\rm Gyr}$, $9.5\,{\rm Gyr}$, $5.2\,{\rm Gyr}$,
and $2.2\,{\rm Gyr}$, respectively.

The mass of the white dwarf is related to the mass of its
progenitor by the initial-to-final mass relation.
Until now, no definitive initial-to-final mass relation
has been established;
however, different estimates exist from
different groups derived from theoretical considerations
and from observational investigations of open clusters; 
see e.g. \citet{weidemann00} and \citet{schroeder01}.
Unfortunately no initial-to-final mass relation for the
halo and the thick disk has been derived yet, so we have to work with what 
is available for the
thin disk and keep in mind that our age 
estimates are crude.
According to \citet{weidemann00}, stars with initial masses
of $1\,\Msolar$, $1.1\,\Msolar$, $1.4\,\Msolar$, and $2\,\Msolar$ would
evolve into white dwarfs with masses of
$0.55\,\Msolar$, $0.555\,\Msolar$, $0.57\,\Msolar$, and $0.6\,\Msolar$, 
respectively.
The initial-to-final mass relation of \citet{schroeder01}, on
the other hand, yields white dwarf masses of
$0.55\,\Msolar$, $0.565\,\Msolar$, $0.605\,\Msolar$, and $0.67\,\Msolar$.

We now estimate how long it takes for a C/O core white dwarf to cool down to
$20\,000\,{\rm K}$, $10\,000\,{\rm K}$, $8\,000\,{\rm K}$, and
$5\,000\,{\rm K}$ using the cooling tracks of \citet{wood95}.
For a $0.5\,\Msolar$ mass white dwarf, the respective cooling times
would be $0.05\,{\rm Gyr}$, $0.5\,{\rm Gyr}$, $0.9\,{\rm Gyr}$,
and $4\,{\rm Gyr}$.
For a $0.6\,\Msolar$ mass white dwarf, the corresponding values are
$0.08\,{\rm Gyr}$, $0.6\,{\rm Gyr}$, $1.1\,{\rm Gyr}$, and
$6\,{\rm Gyr}$.
Hence, only for white dwarfs cooler than 
$8\,000\,{\rm K}$ does
the cooling time contribute significantly to the total age.

All the halo white dwarfs we found have masses less than $0.55\,\Msolar$;
i.e. their
progenitors had a pre-white dwarf life-time of more than $12\,{\rm Gyr}$.
They are all hotter than $14\,000\,{\rm K}$, meaning
they have all cooled less than $0.5\,{\rm Gyr}$.
Due to the large pre-white dwarf lifetime, their total age
is perfectly compatible with halo membership.
It should be noted that the low mass of WD\,0252$-$350 of
only $0.35\,\Msolar$ indicates that it probably does not possess a
CO core but instead a He one.

Now the masses and effective temperatures of the
thick-disk white dwarfs detected in the SPY sample were likewise 
checked. We found that four white dwarfs WD\,0255$-$705,
WD\,0352+052, WD\,1013$-$010, and WD\,1334$-$678 have masses 
which imply ages of less than $7\,{\rm Gyr}$,
which would 
make them too young to belong to the thick disk. 

These four stars are the coolest in our sample of thick-disk
candidates (see Table~\ref{di_par}), with $T_{\rm eff}$ ranging from 8800~K
to 10600~K. \cite{liebert05}
derived the mass
distribution of 348 DA white dwarfs from the PG survey and found that the
average gravities and masses increase with decreasing effective 
temperature for
$T_{\rm eff} < 12000$~K. A similar trend is found in the analysis of more
than 600 DA white dwarfs from the SPY survey (Voss et al., in prep.). 
The physical reason is unknown, but two conjectures have been
published. The high masses inferred from
spectroscopy below $\approx$ 12000~K may actually be due to helium being
brought to the surface by the hydrogen convection zone \citep{bergeron92,
liebert05}. On the other hand, 
\citet{koester05} suggest
that the treatment of non-ideal effects for the level population with the 
Hummer-Mihalas \citep{hummer88} occupation probability mechanism may 
be insufficient for neutral perturbers that become important at lower 
$T_{\rm eff}$. Since these effects are unaccounted for in the model 
atmospheres, 
we may have overestimated the masses
of cool DA white dwarfs ($T_{\rm eff} < 12000$~K).

As a result, the four cool 
white dwarf stars with thick-disk-like kinematics may have a 
lower mass and, therefore, a significantly larger age, one that is perhaps 
even 
consistent with that of the thick disk. Therefore we regard them
as very likely belonging to the thick disk as 
well.

%%%%%%%%%%%%%%%%%%%%%%%%%%%%%%%%%%%%%%%%%%%%%%
%table ha_par
%\input{ha_par_2.tex}

\begin{table}
\caption[]
{Effective temperatures, surface gravities, and masses of the halo white
  dwarfs \label{ha_par}}
\begin{tabular}{llll}
\\
\hline
star & $T_{\rm eff}$ & ${\rm log}~g$ & $M$ \\ 
   &${\rm K}$ & ${\rm cm \,s^{-2}} $  & \Msolar \\
\hline
 HS1527+0614 & 14015 &  7.80 &  0.50 \\
  WD1448+077 & 14459 &  7.66 &  0.44 \\
  WD2351$-$368 & 14567 &  7.81 &  0.51 \\
  WD0252$-$350 & 17056 &  7.42 &  0.35 \\
  WD2359$-$324 & 23267 &  7.65 &  0.47 \\
  WD1524$-$749 & 23414 &  7.61 &  0.45 \\
 HE0201$-$0513 & 24604 &  7.67 &  0.48 \\
\hline
\end{tabular}
\end{table}

%table di_par
%\input{di_par_2.tex}
%%%%%%%%%%%%%%%%%%%%%%%%%%%%%%%%%%%%%%%%%%%%

\begin{table}
\caption[]
{Effective temperatures, surface gravities, and masses of the thick-disk white 
dwarfs. The four coolest stars have higher masses than the rest;
however, the masses of the former may have been overestimated (see text).
\label{di_par}}
\begin{tabular}{llll}
\\
\hline
star & $T_{\rm eff}$ & ${\rm log}~g$ & $M$ \\ 
   &${\rm K}$ & ${\rm cm \,s^{-2}} $  & \Msolar \\
\hline
  WD1013$-$010 &  8786 &  8.19 &  0.71 \\
  WD1334$-$678 &  8958 &  8.11 &  0.66 \\
  WD0352+052 & 10234 &  8.00 &  0.60 \\
  WD0255$-$705 & 10574 &  8.09 &  0.65 \\
\hline
WD1716+020 & 12795 &  7.66 &  0.43 \\
  WD2029+183 & 12976 &  7.73 &  0.47 \\
  WD0204$-$233 & 13176 &  7.75 &  0.47 \\
  WD1952$-$206 & 13742 &  7.78 &  0.49 \\
  WD0732$-$427 & 14070 &  7.96 &  0.58 \\
  WD1327$-$083 & 14141 &  7.79 &  0.50 \\
  WD1614$-$128 & 15313 &  7.74 &  0.48 \\
 HS1432+1441 & 15414 &  7.77 &  0.49 \\
 HE0508$-$2343 & 15835 &  7.71 &  0.47 \\
 HE1124+0144 & 15876 &  7.68 &  0.45 \\
  WD1426$-$276 & 17526 &  7.67 &  0.45 \\
  WD1834$-$781 & 17564 &  7.76 &  0.49 \\
  WD2350$-$083 & 17966 &  7.76 &  0.49 \\
  WD1323$-$514 & 18604 &  7.71 &  0.47 \\
  WD1507+021 & 19384 &  7.79 &  0.51 \\
 HE0452$-$3444 & 20035 &  7.82 &  0.53 \\
  WD1152$-$287 & 20185 &  7.64 &  0.45 \\
  WD1410+168 & 20757 &  7.74 &  0.49 \\
  WD2322$-$181 & 21478 &  7.88 &  0.56 \\
 HE0416$-$1034 & 23809 &  7.88 &  0.57 \\
 HS1338+0807 & 25057 &  7.73 &  0.50 \\
 HE0409$-$5154 & 26439 &  7.75 &  0.52 \\
 HS0820+2503 & 33330 &  7.69 &  0.51 \\
\hline
\end{tabular}
\end{table}

An alternative explanation for the four cool DA stars discussed
above having 
gained thick-disk-like
orbits could be that
they might be run-away stars that were born in a binary system in the
thin disk and were thereafter ejected from it.
Two ejection mechanisms have been suggested. The first one implies a close
binary system in which the primary undergoes a supernova explosion and
releases the secondary at high velocity \citep{davies02}.
This study showed that, indeed, a large fraction of such
binaries are broken up when the primary explodes as a supernova.
A large number of the secondaries receive kick
velocities of $100 \mbox{-}200\,{\rm km s^{-1}}$ and travel on Galactic
orbits similar to those of thick-disk stars.
Thus a population of white dwarfs originating in
the thin disk may contribute significantly to the observed
population of high-velocity white dwarfs.

Another possibility for explaining young white dwarfs with thick-disk-like 
kinematics was proposed by \citet{kroupa02}, who suggests
a scenario for the thickening of galactic disks
through clustered star formation. Massive star clusters may add kinematically
hot components to galactic field populations.

As their masses may be overestimated, we think it is not required to
invoke such run-away scenarios to explain the origin of the four cool white 
dwarfs discussed above. A more natural explanation would be that we have simply
underestimated their ages. 
 
We therefore classify  
those $23$ white dwarfs where age and
kinematics both indicate a thick-disk membership as bona fide 
thick-disk members. 
In addition, the four cool 
white dwarfs are classified as probable thick-disk stars, i.e. 
all 27 stars are retained as thick-disk members. 
This leaves us with a fraction of $2\%$ halo and
$7\%$ thick-disk white dwarfs.

%%%%%%%%%%%%%%%%%%%%%%%%%%%%%%%%%%%%%%%%%%%%%%%%%%%%%%%%%%%%%%%%
\section{Discussion\label{dis}}
We have refined 
and sharpened 
the population classification scheme
developed in Paper I and applied it to a kinematical
analysis of a sample of $398$ DA white dwarfs from the SPY project.
Combining three kinematic criteria, i.e. the position in the $U$\/-$V$-diagram,
the position in the $J_Z$-$e$-diagram, and the Galactic orbit with age 
estimates, we found seven halo and $23$ thick-disk members.

To be able to discuss the kinematic parameters of the three different
populations white dwarfs, we calculated the mean value and standard
deviation of the three velocity components. Of interest are the
asymmetric drift ($V_{\rm lag}=220{\rm km~s^{-1}}-<V>$) for the thick-disk
white dwarfs and the velocity dispersions of the white dwarfs of all
three populations (Tables~5--7).
For comparison, the corresponding values derived by \citet{chiba00}
and \citet{soubiran03} for main-sequence stars are also shown.

The velocity dispersions that were found for the thin-disk
white dwarfs are compatible with the ones of \citet{soubiran03}.
The same is the case for the asymmetric drift and the velocity
dispersions of the thick disk.
Here agreement with the results of \citet{soubiran03} is much
better than with the earlier results of \citet{chiba00}.
There, $\sigma_{\rm U}$ and $\sigma_{\rm V}$ of the halo white dwarfs
are similar to the values of \citet{chiba00}, while
our $\sigma_{\rm W}$ is much smaller.
This is probably due to the fact that our local sample
does not extend as far in the $Z$-direction as the
sample of \citet{chiba00} does.
Also with only seven halo white dwarfs, we have to account for
small number statistics.
In general, the kinematic parameters
of the white dwarfs of the three different populations
do not differ much from those of the main-sequence samples.
\begin{table}
\caption[]
{Standard deviation of $U$, $V$,
$W$ for the $361$ SPY thin-disk
white dwarfs, $\sigma_{\rm U}$, $\sigma_{\rm V}$, and $\sigma_{\rm W}$
from \citet{soubiran03} are shown for comparison \label{uvw_duetab}}
\begin{tabular}{lccc}
\hline
  &$\sigma_{\rm U}$ & $\sigma_{\rm V}$ & $\sigma_{\rm W}$\\
  &${\rm km~s^{-1}}$ & ${\rm km~s^{-1}}$ & ${\rm km~s^{-1}}$\\
\hline
Thin-disk WDs &      &      &    \\
(our sample)        & $34$ & $24$ & $18$\\
\hline
Thin-disk stars &             &      &\\
(Soubiran et al.)& $39$ & $20$ & $20$\\
\hline
\end{tabular}
\end{table}
%%%%%%%%%%%%%%%%%%%%%%%%%%%%%%%%%%%%%%%%%%%%%%%%%%%%%%%%%%%%%%%%%%%%%%%%%%

\begin{table}
\caption[]
{Asymmetric drift $V_{\rm lag}$ and standard deviation of $U$, $V$,
$W$ for the $27$ SPY thick-disk
white dwarfs, $V_{\rm lag}$, $\sigma_{\rm U}$, $\sigma_{\rm V}$, 
and $\sigma_{\rm W}$
from \citet{soubiran03}, and \citet{chiba00} are shown for 
comparison \label{uvw_ditab}}
\begin{tabular}{lcccc}
\hline
 &$V_{\rm lag}$ & $\sigma_{\rm U}$ & $\sigma_{\rm V}$ & $\sigma_{\rm W}$\\
 &${\rm km~s^{-1}}$ & ${\rm km~s^{-1}}$ & ${\rm km~s^{-1}}$ & ${\rm km~s^{-1}}$\\
\hline
Thick-disk WDs &     &      & \\
(our sample)& $-51$ & $79$ & $36$ & $46$\\
\hline
Thick-disk stars        &       &      &      &\\
(Soubiran et al.) & $-51$ & $63$ & $39$ & $39$\\
\hline
Thick-disk stars     &       &      &      &\\
(Chiba \& Beers) & $-20$ & $46$ & $50$ & $35$\\
\hline
\end{tabular}
\end{table}
%%%%%%%%%%%%%%%%%%%%%%%%%%%%%%%%%%%%%%%%%%%%%%%%%%%%%%%%%%%%%%%%%%%%%%%%%%

\begin{table}
\caption[]
{Standard deviation of $U$, $V$,
$W$ for the seven SPY halo
white dwarfs, $V_{\rm lag}$, $\sigma_{\rm U}$, $\sigma_{\rm V}$, and 
$\sigma_{\rm W}$
from \citet{chiba00} shown for comparison \label{uvw_hatab}}
\begin{tabular}{lccc}
\hline
 & $\sigma_{\rm U}$ & $\sigma_{\rm V}$ & $\sigma_{\rm W}$\\
 & ${\rm km~s^{-1}}$ & ${\rm km~s^{-1}}$ & ${\rm km~s^{-1}}$\\
\hline
Halo white dwarfs   & & &\\ 
(our sample)& $138$ & $95$ & $47$\\
\hline
Halo stars  & & &\\
(Chiba \& Beers) &  $141$ & $106$ & $94$\\
\hline
\end{tabular}
\end{table}
%%%%%%%%%%%%%%%%%%%%%%%%%%%%%%%%%%%%%%%%%%%%%%%%%%%%%%%%%%%%%%%%%%%%%%%%%%

We found seven halo white dwarfs in our sample, which
corresponds to a fraction of $2\%$.
In Paper I we found $4\%$ halo white dwarfs,
a deviation possibly due to small number statistics or to
a target selection effect.
In our first paper, we analysed stars from the early phase of
SPY. This sample contained a relatively large fraction of white dwarfs
detected in proper motion surveys 
\citep[][and references therein]{Luyten79, giclas78}. Therefore an
over-representation of white dwarfs with high tangential velocities
is not unexpected.

Our value is lower than the one derived by
\citet{sion88},
who identified about $5$\% of their sample as halo white dwarfs.
\citet{liebert89}, on the other hand, obtained a percentage of
14\% halo white dwarfs
by classifying all stars that exceed a certain value of tangential velocity
as halo members.
When comparing those samples with ours, it has to be kept in mind that
our selection criteria are 
sharper and allow us to separate thick-disk
from halo stars. It is likely that a fraction of the white dwarfs classified 
as halo stars by
\citet{sion88} and \citet{liebert89} actually belong to the thick disk.
Furthermore, both samples suffer from the lack of radial velocity 
information.

It is difficult to compare our sample to the one of \citet{oppenheimer01},
because the inhomogeneous sky coverage of SPY does not allow us to
calculate a space density for halo white dwarfs.
It has to be taken into account that our sample is a magnitude
limited sample and thus biased towards
high temperatures \citep[mean temperature of $21\,000\,\mathrm{K}$; see
also discussion in][]{schroeder04}, 
whereas \citet{oppenheimer01} analyse much cooler white
dwarfs.

Classically, halo white dwarfs are supposed to be cool stars
that originated from high mass progenitors.
The main contribution to the total ages of these white dwarfs is
the cooling time.
This work demonstrates that another
class of hot, low-mass halo white dwarfs exists
with low-mass progenitors that only recently have become white dwarfs so
have not had much time to cool down.
This makes this SPY sample
complement to samples that focus on cool halo white dwarfs.

There are $27$ SPY white dwarfs classified as thick-disk members 
out of which 
four are too cool to allow reliable ages to be derived.
This corresponds to a local fraction of thick-disk white dwarfs of
$7\%$ or $6\%$, if we reject the four cool stars. 
These values are somewhat lower than the $11\%$ found 
by \citet{silvestri02} but are much smaller than that of 
Fuhrmann (2000)\footnote{http://www.xray.mpe.mpg.de/fuhrmann/.},
who predicted a fraction of $17$\% thick-disk white dwarfs.
The differences are possibly caused by the temperature bias mentioned above.
An over-representation of white dwarfs compared to
low mass main-sequence stars, which would require a truncated initial
mass function as suggested by \citet{favata97}, has not been found.

The question of whether thick-disk white dwarfs contribute
significantly to the total mass of the Galaxy is very important for clarifying 
the dark matter problem.
This contribution can be estimated from the results derived above.
To derive the densities of thin-disk and thick-disk white dwarfs,
we used the $1/V_{\rm max}$ method \citep{schmidt68}.
The mass density of thick-disk over thin-disk white dwarfs
${M_{\rm thick}\over M_{\rm thin}}$ was calculated as described
in Paper I.
For the thick disk we adopted the values of
\citet{ojha01}, scale length $l_{\rm 0,thick}=3.7\,{\rm kpc}$, and
tried two extreme values of the
scale height, $h_{\rm 0,thick}=0.8\,{\rm kpc}$ \citep{ojha99} and
$h_{\rm 0,thick}=1.3\,{\rm kpc}$ \citep{chen97}.
For the thin disk, we assumed
$l_{\rm 0,thin}=2.8\,{\rm kpc}$ \citep{ojha01} and
$h_{\rm 0,thin}=0.25\,{\rm kpc}$, in between the values of
\citet{kroupa92} and \citet{haywood97}.
We found ${M_{\rm thick}\over M_{\rm thin}}=0.12 \pm 0.36$ and
${M_{\rm thick}\over M_{\rm thin}}=0.19 \pm 0.57$ for thick-disk scale 
heights of $0.8\,{\rm kpc}$
and $1.3\,{\rm kpc}$, respectively.
Accordingly, upper limits for ${M_{\rm thick}\over M_{\rm thin}}$ are 
$0.48$ and $0.76$, respectively.
Of course the errors are huge because of the poor statistics of the
relatively small thick-disk sample.
Nevertheless, it can be concluded
that the total mass of thick-disk white dwarfs is less than $48\%$
($76\%$) of the total mass of thin-disk white dwarfs.
Therefore the mass contribution of the thick-disk white dwarfs
must not  be neglected, but it is not sufficient to account for the
missing dark matter.

%%%%%%%%%%%%%%%%%%%%%%%%%%%%%%%%%%%%%%%%%%%%%%%%%%%%%%%%%%%%%%%%%%%%%%%%%
%%%%%%%%%%%%%%%%%%%%%%%%%%%%%%%%%%%%%%%%%%%%%%%%%%%%%%%%%%%%%%%%%%%%%%%%%%%%%%
%%%%%%%%%%%%%%%%%%%%%%%%%%%%%%%%%%%%%%%%%%%%%%%%%%%%%%%%%%%%%%%%%%%%%%%%%%%%
\section{Conclusions\label{con}}
We have demonstrated how a combination of sophisticated kinematic analysis
tools can distinguish halo, thick-disk, and thin-disk white
dwarfs. We identified a fraction of 2\% halo and 7\% thick-disk white
dwarfs.
Most of our thick-disk and halo white dwarfs
are hot and possess low masses.
Our results suggest that the mass present in halo and thick-disk
white dwarfs is not sufficient for explaining the missing mass of the Galaxy.
But to draw definite conclusions, more data are needed.
Our goal is to extend this kinematic analysis to all
$1\,000$ 
degenerate stars
from the SPY project,
in order to have a large data base for
deciding on the population membership of white dwarfs
and their implications for the mass and evolution of the Galaxy.
%%%%%%%%%%%%%%%%%%%%%%%%%%%%%%%%%%%%%%%%%%%%%%%%%%%%%%%%%%%%%%%%%%%%%%
\acknowledgements { We thank D. Koester for providing us with the
results of his spectral analysis prior to publication and B. Voss for prolific
discussions. 
E.-M. P. acknowledges support by the Deutsche
Forschungsgemeinschaft (grant Na\,365/2-1) and is grateful to the
Studienstiftung des Deutschen Volkes for a grant.  M. Altmann
acknowledges support from the DLR~50~QD~0102 and from FONDAP~1501~0003.  
R.N.\ is supported by a
PPARC Advanced Fellowship. Thanks go to
J.~Pauli for interesting and fruitful discussions. 
This research has made use of the SIMBAD database,
operated at the CDS, Strasbourg, France 
and of DSS images based on photographic data obtained with the UK Schmidt
Telescope.  
}
%%%%%%%%%%%%%%%%%%%%%%%%%%%%%%%%%%%%%%%%%%%%%%%%%%%%%%%%%%%%%%%%%%%%%%%%
%%%%%%%%%%%%%%%%%%%%%%%%%%%%%%%%%%%%%%%%%%%%%%%%%%%%%%%%%%%%%%%%%
\bibliographystyle{aa}
% \bibliography{paper2}
\bibliography{emp}
%table1
%\clearpage
\begin{table*}[ht]
\caption[]
{Radial velocities (corrected for gravitational redshift), proper motions, and 
spectroscopic distances (with errors) of the $398$ SPY white dwarfs. The 
* indicates stars with only one spectrum\label{inputkin}}
\begin{tabular}{lr@{$\pm$}lr@{$\pm$}lr@{$\pm$}lrr@{$\pm$}l}
\hline
star & \multicolumn{2}{c}{$v_{\rm rad}$} & \multicolumn{2}{r}{$\mu_{\alpha}\,{\rm cos} \delta$} & \multicolumn{2}{c}{$\mu_{\delta}$} & $d$ & \multicolumn{2}{c}{${\rm log}~d$}\\ 
 & \multicolumn{2}{c}{${\rm km\,s^{-1}}$} & \multicolumn{2}{r}{${\rm mas\,yr^{-1}}$} & \multicolumn{2}{r}{${\rm mas\,yr^{-1}}$} & ${\rm pc}$& \multicolumn{2}{c}{${\rm pc}$} \\
\hline
EC\,12489--2750 & \hspace*{.3cm} $ 24.8$ & $17.9$ & \hspace*{.8cm} $  -7$ & $   3$ & \hspace*{.45cm} $ -16$ & $   3$ & $ 431.5$ & $2.63$ & $0.05$ \\
EC\,13123--2523 & \hspace*{.3cm} $ 16.1$ & $ 6.7$ & \hspace*{.8cm} $ -13$ & $   4$ & \hspace*{.45cm} $   9$ & $   7$ & $ 572.0$ & $2.76$ & $0.05$ \\
HE\,0032--2744 & \hspace*{.3cm} $ 30.6$ & $ 4.2$ & \hspace*{.8cm} $  31$ & $   5$ & \hspace*{.45cm} $ -40$ & $   3$ & $ 183.2$ & $2.26$ & $0.04$ \\
HE\,0043--0318 & \hspace*{.3cm} $ 46.3$ & $ 3.3$ & \hspace*{.8cm} $  11$ & $   6$ & \hspace*{.45cm} $ -68$ & $   7$ & $  63.4$ & $1.80$ & $0.04$ \\
HE\,0049--0940 & \hspace*{.3cm} $ 13.2$ & $ 2.8$ & \hspace*{.8cm} $  87$ & $  11$ & \hspace*{.45cm} $  -8$ & $  11$ & $  99.0$ & $2.00$ & $0.04$ \\
HE\,0103--3253 & \hspace*{.3cm} $ 17.1$ & $ 3.6$ & \hspace*{.8cm} $  72$ & $   4$ & \hspace*{.45cm} $  64$ & $   3$ & $  94.1$ & $1.97$ & $0.04$ \\
\hline
\end{tabular}
\end{table*}
\begin{table*}[ht]
\addtocounter{table}{-1}
\caption[]
{Radial velocities (corrected for gravitational redshift), proper motions, 
and spectroscopic distances (with errors) of the $398$ SPY white dwarfs. 
The * indicates stars with only one spectrum, continued from previous page}
\begin{tabular}{lr@{$\pm$}lr@{$\pm$}lr@{$\pm$}lrr@{$\pm$}l}
\hline
star & \multicolumn{2}{c}{$v_{\rm rad}$} & \multicolumn{2}{r}{$\mu_{\alpha}\,{\rm cos} \delta$} & \multicolumn{2}{c}{$\mu_{\delta}$} & $d$ & \multicolumn{2}{c}{${\rm log}~d$}\\ 
 & \multicolumn{2}{c}{${\rm km\,s^{-1}}$} & \multicolumn{2}{r}{${\rm mas\,yr^{-1}}$} & \multicolumn{2}{r}{${\rm mas\,yr^{-1}}$} & ${\rm pc}$& \multicolumn{2}{c}{${\rm pc}$} \\
\hline
HE\,1441--0047 & \hspace*{.3cm} $ -7.3$ & $ 7.0$ & \hspace*{.8cm} $  36$ & $   4$ & \hspace*{.45cm} $  20$ & $   5$ & $ 107.6$ & $2.03$ & $0.04$ \\
HE\,1518--0020 & \hspace*{.3cm} $ 10.9$ & $ 3.7$ & \hspace*{.8cm} $ -97$ & $   5$ & \hspace*{.45cm} $-118$ & $   6$ & $  73.0$ & $1.86$ & $0.04$ \\
HE\,1518--0344 & \hspace*{.3cm} $  2.7$ & $ 5.5$ & \hspace*{.8cm} $ -12$ & $   7$ & \hspace*{.45cm} $ -36$ & $   4$ & $ 203.9$ & $2.31$ & $0.04$ \\
HE\,2133--1332 & \hspace*{.3cm} $-11.9$ & $ 3.2$ & \hspace*{.8cm} $ 259$ & $   7$ & \hspace*{.45cm} $-141$ & $   6$ & $  23.3$ & $1.37$ & $0.04$ \\
HE\,2135--4055 & \hspace*{.3cm} $ 39.6$ & $ 3.8$* & \hspace*{.8cm} $-140$ & $   4$ & \hspace*{.45cm} $  44$ & $   4$ & $  33.2$ & $1.52$ & $0.03$ \\
HE\,2140--1825 & \hspace*{.3cm} $-14.8$ & $ 3.2$ & \hspace*{.8cm} $ -12$ & $   3$ & \hspace*{.45cm} $ -39$ & $   3$ & $  90.3$ & $1.96$ & $0.04$ \\
\hline
\end{tabular}
\end{table*}
\begin{table*}[ht]
\addtocounter{table}{-1}
\caption[]
{Radial velocities (corrected for gravitational redshift), proper motions, and 
spectroscopic distances (with errors) of the $398$ SPY white dwarfs. The* 
indicates stars with only one spectrum, continued from previous page}
\begin{tabular}{lr@{$\pm$}lr@{$\pm$}lr@{$\pm$}lrr@{$\pm$}l}
\hline
star & \multicolumn{2}{c}{$v_{\rm rad}$} & \multicolumn{2}{r}{$\mu_{\alpha}\,{\rm cos} \delta$} & \multicolumn{2}{c}{$\mu_{\delta}$} & $d$ & \multicolumn{2}{c}{${\rm log}~d$}\\ 
 & \multicolumn{2}{c}{${\rm km\,s^{-1}}$} & \multicolumn{2}{r}{${\rm mas\,yr^{-1}}$} & \multicolumn{2}{r}{${\rm mas\,yr^{-1}}$} & ${\rm pc}$& \multicolumn{2}{c}{${\rm pc}$} \\
\hline
MCT\,0105--1634 & \hspace*{.3cm} $  3.2$ & $ 5.9$ & \hspace*{.8cm} $  11$ & $   4$ & \hspace*{.45cm} $ -46$ & $   7$ & $ 315.2$ & $2.50$ & $0.05$ \\
MCT\,0110--1617 & \hspace*{.3cm} $ 27.6$ & $ 8.8$ & \hspace*{.8cm} $   3$ & $  24$ & \hspace*{.45cm} $   6$ & $   7$ & $ 318.3$ & $2.50$ & $0.05$ \\
MCT\,2343--1740 & \hspace*{.3cm} $ -0.0$ & $ 6.8$ & \hspace*{.8cm} $   6$ & $   8$ & \hspace*{.45cm} $  22$ & $  10$ & $ 141.3$ & $2.15$ & $0.04$ \\
MCT\,2345--3940 & \hspace*{.3cm} $-19.8$ & $ 4.2$ & \hspace*{.8cm} $  85$ & $   3$ & \hspace*{.45cm} $  -1$ & $   2$ & $ 124.4$ & $2.09$ & $0.03$ \\
MCT\,2349--3627 & \hspace*{.3cm} $ 15.1$ & $15.6$ & \hspace*{.8cm} $  16$ & $   6$ & \hspace*{.45cm} $ -34$ & $   4$ & $ 400.5$ & $2.60$ & $0.06$ \\
WD\,0000--186 & \hspace*{.3cm} $  2.1$ & $ 4.0$ & \hspace*{.8cm} $ -59$ & $   5$ & \hspace*{.45cm} $ -89$ & $   4$ & $ 110.8$ & $2.04$ & $0.04$ \\
\hline
\end{tabular}
\end{table*}
\begin{table*}[ht]
\addtocounter{table}{-1}
\caption[]
{Radial velocities (corrected for gravitational redshift), proper motions, and 
spectroscopic distances (with errors) of the $398$ SPY white dwarfs. The * 
indicates stars with only one spectrum, continued from previous page}
\begin{tabular}{lr@{$\pm$}lr@{$\pm$}lr@{$\pm$}lrr@{$\pm$}l}
\hline
star & \multicolumn{2}{c}{$v_{\rm rad}$} & \multicolumn{2}{r}{$\mu_{\alpha}\,{\rm cos} \delta$} & \multicolumn{2}{c}{$\mu_{\delta}$} & $d$ & \multicolumn{2}{c}{${\rm log}~d$}\\ 
 & \multicolumn{2}{c}{${\rm km\,s^{-1}}$} & \multicolumn{2}{r}{${\rm mas\,yr^{-1}}$} & \multicolumn{2}{r}{${\rm mas\,yr^{-1}}$} & ${\rm pc}$& \multicolumn{2}{c}{${\rm pc}$} \\
\hline
WD\,0352+096 & \hspace*{.3cm} $ 46.3$ & $ 4.2$ & \hspace*{.8cm} $ 168$ & $   4$ & \hspace*{.45cm} $   4$ & $   4$ & $  38.7$ & $1.59$ & $0.04$ \\
WD\,0408--041 & \hspace*{.3cm} $ -7.6$ & $ 4.2$ & \hspace*{.8cm} $   4$ & $   7$ & \hspace*{.45cm} $-125$ & $   8$ & $  73.3$ & $1.87$ & $0.03$ \\
WD\,0416--550 & \hspace*{.3cm} $ 24.2$ & $ 3.7$ & \hspace*{.8cm} $  47$ & $   8$ & \hspace*{.45cm} $  11$ & $   6$ & $ 226.6$ & $2.36$ & $0.05$ \\
WD\,0431+126 & \hspace*{.3cm} $ 43.5$ & $ 4.0$ & \hspace*{.8cm} $  97$ & $   3$ & \hspace*{.45cm} $ -13$ & $   3$ & $  50.8$ & $1.71$ & $0.04$ \\
WD\,0437+152 & \hspace*{.3cm} $ 10.8$ & $ 3.0$ & \hspace*{.8cm} $ -14$ & $   4$ & \hspace*{.45cm} $ -38$ & $   4$ & $ 153.2$ & $2.19$ & $0.04$ \\
WD\,0455--282 & \hspace*{.3cm} $ 49.2$ & $11.5$ & \hspace*{.8cm} $  56$ & $   2$ & \hspace*{.45cm} $  10$ & $   3$ & $ 115.9$ & $2.06$ & $0.05$ \\
\hline
\end{tabular}
\end{table*}
\begin{table*}[ht]
\addtocounter{table}{-1}
\caption[]
{Radial velocities (corrected for gravitational redshift), proper motions, and 
spectroscopic distances (with errors) of the $398$ SPY white dwarfs. The * 
indicates stars with only one spectrum, continued from previous page}
\begin{tabular}{lr@{$\pm$}lr@{$\pm$}lr@{$\pm$}lrr@{$\pm$}l}
\hline
star & \multicolumn{2}{c}{$v_{\rm rad}$} & \multicolumn{2}{r}{$\mu_{\alpha}\,{\rm cos} \delta$} & \multicolumn{2}{c}{$\mu_{\delta}$} & $d$ & \multicolumn{2}{c}{${\rm log}~d$}\\ 
 & \multicolumn{2}{c}{${\rm km\,s^{-1}}$} & \multicolumn{2}{r}{${\rm mas\,yr^{-1}}$} & \multicolumn{2}{r}{${\rm mas\,yr^{-1}}$} & ${\rm pc}$& \multicolumn{2}{c}{${\rm pc}$} \\
\hline
WD\,1207--157 & \hspace*{.3cm} $-12.9$ & $ 3.8$ & \hspace*{.8cm} $ -72$ & $   6$ & \hspace*{.45cm} $  -2$ & $   8$ & $ 140.2$ & $2.15$ & $0.04$ \\
WD\,1216+036 & \hspace*{.3cm} $ 14.2$ & $ 3.6$ & \hspace*{.8cm} $  12$ & $   5$ & \hspace*{.45cm} $ -82$ & $   4$ & $  89.4$ & $1.95$ & $0.04$ \\
WD\,1220--292 & \hspace*{.3cm} $ -5.8$ & $ 3.5$ & \hspace*{.8cm} $  37$ & $   5$ & \hspace*{.45cm} $ -45$ & $   5$ & $  96.3$ & $1.98$ & $0.04$ \\
WD\,1230--308 & \hspace*{.3cm} $  6.4$ & $ 5.6$ & \hspace*{.8cm} $  32$ & $   2$ & \hspace*{.45cm} $ -20$ & $   2$ & $  93.9$ & $1.97$ & $0.04$ \\
WD\,1231--141 & \hspace*{.3cm} $-17.1$ & $ 3.9$ & \hspace*{.8cm} $ -86$ & $   3$ & \hspace*{.45cm} $ -10$ & $   4$ & $ 105.7$ & $2.02$ & $0.03$ \\
WD\,1233--164 & \hspace*{.3cm} $ 19.7$ & $ 5.3$ & \hspace*{.8cm} $ -91$ & $   6$ & \hspace*{.45cm} $ -56$ & $   5$ & $  74.4$ & $1.87$ & $0.04$ \\
\hline
\end{tabular}
\end{table*}
\begin{table*}[ht]
\addtocounter{table}{-1}
\caption[]
{Radial velocities (corrected for gravitational redshift), proper motions, and 
spectroscopic distances (with errors) of the $398$ SPY white dwarfs. The * 
indicates stars with only one spectrum, continued from previous page}
\begin{tabular}{lr@{$\pm$}lr@{$\pm$}lr@{$\pm$}lrr@{$\pm$}l}
\hline
star & \multicolumn{2}{c}{$v_{\rm rad}$} & \multicolumn{2}{r}{$\mu_{\alpha}\,{\rm cos} \delta$} & \multicolumn{2}{c}{$\mu_{\delta}$} & $d$ & \multicolumn{2}{c}{${\rm log}~d$}\\ 
 & \multicolumn{2}{c}{${\rm km\,s^{-1}}$} & \multicolumn{2}{r}{${\rm mas\,yr^{-1}}$} & \multicolumn{2}{r}{${\rm mas\,yr^{-1}}$} & ${\rm pc}$& \multicolumn{2}{c}{${\rm pc}$} \\
\hline
WD\,1716+020 & \hspace*{.3cm} $-37.5$ & $ 3.1$* & \hspace*{.8cm} $-424$ & $   4$ & \hspace*{.45cm} $-300$ & $   5$ & $  44.0$ & $1.64$ & $0.04$ \\
WD\,1736+052 & \hspace*{.3cm} $  8.5$ & $ 6.0$ & \hspace*{.8cm} $ -43$ & $  10$ & \hspace*{.45cm} $-261$ & $  10$ & $  39.9$ & $1.60$ & $0.04$ \\
WD\,1826--045 & \hspace*{.3cm} $-35.0$ & $ 4.4$* & \hspace*{.8cm} $  11$ & $   7$ & \hspace*{.45cm} $-298$ & $   4$ & $  24.7$ & $1.39$ & $0.04$ \\
WD\,1827--106 & \hspace*{.3cm} $-54.7$ & $ 3.6$* & \hspace*{.8cm} $ 254$ & $  17$ & \hspace*{.45cm} $-306$ & $  17$ & $  45.6$ & $1.66$ & $0.04$ \\
WD\,1834--781 & \hspace*{.3cm} $ 71.1$ & $ 3.5$ & \hspace*{.8cm} $ 130$ & $  16$ & \hspace*{.45cm} $-313$ & $  16$ & $  92.7$ & $1.97$ & $0.03$ \\
WD\,1840+042 & \hspace*{.3cm} $-38.9$ & $ 4.9$* & \hspace*{.8cm} $-112$ & $  16$ & \hspace*{.45cm} $  62$ & $   5$ & $  25.2$ & $1.40$ & $0.04$ \\
\hline
\end{tabular}
\end{table*}
\begin{table*}[ht]
\addtocounter{table}{-1}
\caption[]
{Radial velocities (corrected for gravitational redshift), proper motions, and 
spectroscopic distances (with errors) of the $398$ SPY white dwarfs. The * 
indicates stars with only one spectrum, continued from previous page}
\begin{tabular}{lr@{$\pm$}lr@{$\pm$}lr@{$\pm$}lrr@{$\pm$}l}
\hline
star & \multicolumn{2}{c}{$v_{\rm rad}$} & \multicolumn{2}{r}{$\mu_{\alpha}\,{\rm cos} \delta$} & \multicolumn{2}{c}{$\mu_{\delta}$} & $d$ & \multicolumn{2}{c}{${\rm log}~d$}\\ 
 & \multicolumn{2}{c}{${\rm km\,s^{-1}}$} & \multicolumn{2}{r}{${\rm mas\,yr^{-1}}$} & \multicolumn{2}{r}{${\rm mas\,yr^{-1}}$} & ${\rm pc}$& \multicolumn{2}{c}{${\rm pc}$} \\
\hline
WD\,2347--192 & \hspace*{.3cm} $  7.9$ & $ 5.2$ & \hspace*{.8cm} $ -53$ & $   7$ & \hspace*{.45cm} $ -13$ & $   4$ & $ 167.3$ & $2.22$ & $0.04$ \\
WD\,2348--244 & \hspace*{.3cm} $ 34.0$ & $ 5.2$ & \hspace*{.8cm} $-105$ & $   6$ & \hspace*{.45cm} $-229$ & $   3$ & $  49.5$ & $1.69$ & $0.04$ \\
WD\,2349--283 & \hspace*{.3cm} $ 14.8$ & $ 3.4$ & \hspace*{.8cm} $  36$ & $   2$ & \hspace*{.45cm} $ -57$ & $   2$ & $  97.4$ & $1.99$ & $0.04$ \\
WD\,2350--083 & \hspace*{.3cm} $ 37.3$ & $ 3.6$ & \hspace*{.8cm} $ 167$ & $  30$ & \hspace*{.45cm} $  11$ & $  16$ & $ 132.2$ & $2.12$ & $0.04$ \\
WD\,2350--248 & \hspace*{.3cm} $ 17.1$ & $ 6.2$ & \hspace*{.8cm} $ -19$ & $   2$ & \hspace*{.45cm} $  -4$ & $   2$ & $ 100.6$ & $2.00$ & $0.04$ \\
WD\,2351--368 & \hspace*{.3cm} $-28.8$ & $ 5.0$ & \hspace*{.8cm} $  33$ & $  17$ & \hspace*{.45cm} $-680$ & $  17$ & $  62.4$ & $1.80$ & $0.04$ \\
WD\,2354--151 & \hspace*{.3cm} $ 11.7$ & $ 5.0$ & \hspace*{.8cm} $  -5$ & $   3$ & \hspace*{.45cm} $ -18$ & $   3$ & $ 278.2$ & $2.44$ & $0.04$ \\
WD\,2359--324 & \hspace*{.3cm} $ 64.1$ & $ 4.2$ & \hspace*{.8cm} $ 266$ & $   4$ & \hspace*{.45cm} $ -89$ & $   4$ & $ 161.9$ & $2.21$ & $0.04$ \\
WD\,2359--434 & \hspace*{.3cm} $-58.8$ & $10.8$ & \hspace*{.8cm} $ 719$ & $  53$ & \hspace*{.45cm} $-724$ & $  53$ & $   6.1$ & $0.79$ & $0.04$ \\
\hline
\end{tabular}
\end{table*}

\clearpage
%table2
\begin{table*}[h!]
\caption[]
{Kinematic parameters of the $398$ SPY white dwarfs\label{kinpar2}}
\begin{tabular}{lr@{$\pm$}lr@{$\pm$}lr@{$\pm$}lr@{$\pm$}lr@{$\pm$}l}
\hline
star & \multicolumn{2}{c}{$e$} & \multicolumn{2}{c}{$J_z$} & \multicolumn{2}{c}{$U$} & \multicolumn{2}{c}{$V$} & \multicolumn{2}{c}{$W$} \\ 
 & \multicolumn{2}{c}{}& \multicolumn{2}{c}{${\rm kpc\,km\,s^{-1}}$} & \multicolumn{2}{c}{${\rm km\,s^{-1}}$} & \multicolumn{2}{c}{${\rm km\,s^{-1}}$} & \multicolumn{2}{c}{${\rm km\,s^{-1}}$} \\
\hline
EC\,12489$-$2750 & $0.13$ & $0.05$ & \hspace*{.4cm} $ 1554$ & $   93$ & $20.0$ & $ 8.4$ & $199.8$ & $11.6$ & $-4.8$ & $10.2$ \\
EC\,13123$-$2523 & $0.06$ & $0.04$ & \hspace*{.4cm} $ 1726$ & $   84$ & $-16.2$ & $ 9.4$ & $223.2$ & $10.7$ & $39.3$ & $14.6$ \\
HE\,0032$-$2744 & $0.10$ & $0.02$ & \hspace*{.4cm} $ 1577$ & $   31$ & $ 6.8$ & $ 3.1$ & $197.3$ & $ 3.9$ & $-24.5$ & $ 3.4$ \\
HE\,0043$-$0318 & $0.10$ & $0.01$ & \hspace*{.4cm} $ 1913$ & $   17$ & $ 7.9$ & $ 1.7$ & $238.8$ & $ 2.1$ & $-43.0$ & $ 2.7$ \\
HE\,0049$-$0940 & $0.08$ & $0.02$ & \hspace*{.4cm} $ 1748$ & $   36$ & $-24.4$ & $ 4.7$ & $218.1$ & $ 4.5$ & $-6.1$ & $ 2.5$ \\
HE\,0103$-$3253 & $0.15$ & $0.01$ & \hspace*{.4cm} $ 1945$ & $   10$ & $-32.5$ & $ 3.1$ & $243.1$ & $ 1.2$ & $-10.6$ & $ 2.8$ \\
\hline
\end{tabular}
\end{table*}
%%%%%%%%%%%%%%%%%%%%%%%%%%%%%%%%%%%%%%%%%%%%%%%%%%%%%%%%%%
\begin{table*}[h!]
\addtocounter{table}{-1}
\caption[]
{Kinematic parameters of the $398$ SPY white dwarfs\\
continued from previous page}
\begin{tabular}{lr@{$\pm$}lr@{$\pm$}lr@{$\pm$}lr@{$\pm$}lr@{$\pm$}l}
\hline
star & \multicolumn{2}{c}{$e$} & \multicolumn{2}{c}{$J_z$} & \multicolumn{2}{c}{$U$} & \multicolumn{2}{c}{$V$} & \multicolumn{2}{c}{$W$} \\ 
 & \multicolumn{2}{c}{}& \multicolumn{2}{c}{${\rm kpc\,km\,s^{-1}}$} & \multicolumn{2}{c}{${\rm km\,s^{-1}}$} & \multicolumn{2}{c}{${\rm km\,s^{-1}}$} & \multicolumn{2}{c}{${\rm km\,s^{-1}}$} \\
\hline
HE\,1441$-$0047 & $0.18$ & $0.01$ & \hspace*{.4cm} $ 2063$ & $   18$ & $12.2$ & $ 3.9$ & $260.1$ & $ 2.4$ & $-1.2$ & $ 4.5$ \\
HE\,1518$-$0020 & $0.16$ & $0.02$ & \hspace*{.4cm} $ 1490$ & $   31$ & $20.6$ & $ 2.5$ & $187.4$ & $ 3.8$ & $15.2$ & $ 2.3$ \\
HE\,1518$-$0344 & $0.09$ & $0.02$ & \hspace*{.4cm} $ 1618$ & $   38$ & $21.1$ & $ 4.4$ & $206.2$ & $ 4.7$ & $-1.7$ & $ 4.5$ \\
HE\,2133$-$1332 & $0.04$ & $0.01$ & \hspace*{.4cm} $ 1747$ & $   14$ & $-12.3$ & $ 1.9$ & $218.7$ & $ 1.7$ & $-8.6$ & $ 2.5$ \\
HE\,2135$-$4055 & $0.21$ & $0.01$ & \hspace*{.4cm} $ 1988$ & $    6$ & $52.3$ & $ 2.4$ & $249.2$ & $ 0.7$ & $-7.8$ & $ 2.5$ \\
HE\,2140$-$1825 & $0.03$ & $0.01$ & \hspace*{.4cm} $ 1745$ & $   15$ & $10.9$ & $ 1.9$ & $219.5$ & $ 1.8$ & $17.5$ & $ 2.0$ \\
\hline
\end{tabular}
\end{table*}
%%%%%%%%%%%%%%%%%%%%%%%%%%%%%%%%%%%%%%%%%%%%%%%%%%%%%%%%%%
\begin{table*}[h!]
\addtocounter{table}{-1}
\caption[]
{Kinematic parameters of the $398$ SPY white dwarfs\\
continued from previous page}
\begin{tabular}{lr@{$\pm$}lr@{$\pm$}lr@{$\pm$}lr@{$\pm$}lr@{$\pm$}l}
\hline
star & \multicolumn{2}{c}{$e$} & \multicolumn{2}{c}{$J_z$} & \multicolumn{2}{c}{$U$} & \multicolumn{2}{c}{$V$} & \multicolumn{2}{c}{$W$} \\ 
 & \multicolumn{2}{c}{}& \multicolumn{2}{c}{${\rm kpc\,km\,s^{-1}}$} & \multicolumn{2}{c}{${\rm km\,s^{-1}}$} & \multicolumn{2}{c}{${\rm km\,s^{-1}}$} & \multicolumn{2}{c}{${\rm km\,s^{-1}}$} \\
\hline
MCT\,0105$-$1634 & $0.22$ & $0.04$ & \hspace*{.4cm} $ 1410$ & $   75$ & $34.4$ & $ 6.7$ & $175.0$ & $ 9.4$ & $-7.8$ & $ 5.0$ \\
MCT\,0110$-$1617 & $0.16$ & $0.07$ & \hspace*{.4cm} $ 2003$ & $  148$ & $-3.7$ & $23.9$ & $248.6$ & $18.2$ & $-17.5$ & $ 7.7$ \\
MCT\,2343$-$1740 & $0.14$ & $0.02$ & \hspace*{.4cm} $ 2003$ & $   41$ & $-0.1$ & $ 4.6$ & $251.0$ & $ 5.2$ & $ 9.2$ & $ 5.5$ \\
MCT\,2345$-$3940 & $0.13$ & $0.01$ & \hspace*{.4cm} $ 1753$ & $   14$ & $-39.6$ & $ 3.4$ & $220.1$ & $ 1.7$ & $14.4$ & $ 3.4$ \\
MCT\,2349$-$3627 & $0.24$ & $0.05$ & \hspace*{.4cm} $ 1319$ & $   89$ & $13.4$ & $ 9.1$ & $167.1$ & $11.0$ & $-4.5$ & $12.1$ \\
WD\,0000$-$186 & $0.19$ & $0.01$ & \hspace*{.4cm} $ 1710$ & $   21$ & $60.3$ & $ 4.1$ & $213.9$ & $ 2.6$ & $ 3.9$ & $ 3.1$ \\
\hline
\end{tabular}
\end{table*}
%%%%%%%%%%%%%%%%%%%%%%%%%%%%%%%%%%%%%%%%%%%%%%%%%%%%%%%%%%
\begin{table*}[h!]
\addtocounter{table}{-1}
\caption[]
{Kinematic parameters of the $398$ SPY white dwarfs\\
continued from previous page}
\begin{tabular}{lr@{$\pm$}lr@{$\pm$}lr@{$\pm$}lr@{$\pm$}lr@{$\pm$}l}
\hline
star & \multicolumn{2}{c}{$e$} & \multicolumn{2}{c}{$J_z$} & \multicolumn{2}{c}{$U$} & \multicolumn{2}{c}{$V$} & \multicolumn{2}{c}{$W$} \\ 
 & \multicolumn{2}{c}{}& \multicolumn{2}{c}{${\rm kpc\,km\,s^{-1}}$} & \multicolumn{2}{c}{${\rm km\,s^{-1}}$} & \multicolumn{2}{c}{${\rm km\,s^{-1}}$} & \multicolumn{2}{c}{${\rm km\,s^{-1}}$} \\
\hline
WD\,0352+096 & $0.13$ & $0.01$ & \hspace*{.4cm} $ 1771$ & $   12$ & $-42.0$ & $ 3.2$ & $220.5$ & $ 1.5$ & $ 3.1$ & $ 2.4$ \\
WD\,0408$-$041 & $0.13$ & $0.01$ & \hspace*{.4cm} $ 1682$ & $   23$ & $38.6$ & $ 3.3$ & $208.8$ & $ 2.9$ & $-4.9$ & $ 2.7$ \\
WD\,0416$-$550 & $0.13$ & $0.03$ & \hspace*{.4cm} $ 1553$ & $   46$ & $-19.7$ & $ 6.1$ & $193.3$ & $ 5.9$ & $23.9$ & $ 6.2$ \\
WD\,0431+126 & $0.11$ & $0.01$ & \hspace*{.4cm} $ 1773$ & $   10$ & $-35.3$ & $ 3.1$ & $220.3$ & $ 1.3$ & $ 6.1$ & $ 1.7$ \\
WD\,0437+152 & $0.04$ & $0.01$ & \hspace*{.4cm} $ 1825$ & $   20$ & $ 9.0$ & $ 2.6$ & $224.1$ & $ 2.5$ & $-19.5$ & $ 2.9$ \\
WD\,0455$-$282 & $0.14$ & $0.03$ & \hspace*{.4cm} $ 1568$ & $   47$ & $-26.1$ & $ 5.1$ & $194.3$ & $ 6.0$ & $ 3.9$ & $ 6.0$ \\
\hline
\end{tabular}
\end{table*}
%%%%%%%%%%%%%%%%%%%%%%%%%%%%%%%%%%%%%%%%%%%%%%%%%%%%%%%%%%
\begin{table*}[h!]
\addtocounter{table}{-1}
\caption[]
{Kinematic parameters of the $398$ SPY white dwarfs\\
continued from previous page}
\begin{tabular}{lr@{$\pm$}lr@{$\pm$}lr@{$\pm$}lr@{$\pm$}lr@{$\pm$}l}
\hline
star & \multicolumn{2}{c}{$e$} & \multicolumn{2}{c}{$J_z$} & \multicolumn{2}{c}{$U$} & \multicolumn{2}{c}{$V$} & \multicolumn{2}{c}{$W$} \\ 
 & \multicolumn{2}{c}{}& \multicolumn{2}{c}{${\rm kpc\,km\,s^{-1}}$} & \multicolumn{2}{c}{${\rm km\,s^{-1}}$} & \multicolumn{2}{c}{${\rm km\,s^{-1}}$} & \multicolumn{2}{c}{${\rm km\,s^{-1}}$} \\
\hline
WD\,1207$-$157 & $0.11$ & $0.01$ & \hspace*{.4cm} $ 1801$ & $   32$ & $-34.1$ & $ 4.4$ & $225.6$ & $ 4.0$ & $-9.9$ & $ 3.8$ \\
WD\,1216+036 & $0.12$ & $0.01$ & \hspace*{.4cm} $ 1675$ & $   19$ & $32.8$ & $ 2.2$ & $209.8$ & $ 2.4$ & $ 7.0$ & $ 2.9$ \\
WD\,1220$-$292 & $0.15$ & $0.01$ & \hspace*{.4cm} $ 1940$ & $   20$ & $29.6$ & $ 2.4$ & $243.8$ & $ 2.5$ & $-11.2$ & $ 2.5$ \\
WD\,1230$-$308 & $0.13$ & $0.01$ & \hspace*{.4cm} $ 1901$ & $   26$ & $27.5$ & $ 2.2$ & $239.1$ & $ 3.3$ & $ 4.3$ & $ 2.4$ \\
WD\,1231$-$141 & $0.09$ & $0.01$ & \hspace*{.4cm} $ 1795$ & $   22$ & $-30.1$ & $ 2.9$ & $225.0$ & $ 2.8$ & $-11.5$ & $ 2.7$ \\
WD\,1233$-$164 & $0.09$ & $0.02$ & \hspace*{.4cm} $ 1591$ & $   28$ & $-3.2$ & $ 2.5$ & $199.4$ & $ 3.4$ & $ 6.3$ & $ 3.4$ \\
\hline
\end{tabular}
\end{table*}
%%%%%%%%%%%%%%%%%%%%%%%%%%%%%%%%%%%%%%%%%%%%%%%%%%%%%%%%%%
\begin{table*}[h!]
\addtocounter{table}{-1}
\caption[]
{Kinematic parameters of the $398$ SPY white dwarfs\\
continued from previous page}
\begin{tabular}{lr@{$\pm$}lr@{$\pm$}lr@{$\pm$}lr@{$\pm$}lr@{$\pm$}l}
\hline
star & \multicolumn{2}{c}{$e$} & \multicolumn{2}{c}{$J_z$} & \multicolumn{2}{c}{$U$} & \multicolumn{2}{c}{$V$} & \multicolumn{2}{c}{$W$} \\ 
 & \multicolumn{2}{c}{}& \multicolumn{2}{c}{${\rm kpc\,km\,s^{-1}}$} & \multicolumn{2}{c}{${\rm km\,s^{-1}}$} & \multicolumn{2}{c}{${\rm km\,s^{-1}}$} & \multicolumn{2}{c}{${\rm km\,s^{-1}}$} \\
\hline
WD\,1716+020 & $0.40$ & $0.03$ & \hspace*{.4cm} $ 1033$ & $   56$ & $ 1.0$ & $ 2.7$ & $129.7$ & $ 7.0$ & $38.3$ & $ 3.3$ \\
WD\,1736+052 & $0.16$ & $0.01$ & \hspace*{.4cm} $ 1630$ & $   30$ & $44.5$ & $ 4.5$ & $204.5$ & $ 3.7$ & $-4.3$ & $ 2.3$ \\
WD\,1826$-$045 & $0.10$ & $0.01$ & \hspace*{.4cm} $ 1575$ & $   21$ & $-7.1$ & $ 3.4$ & $197.5$ & $ 2.6$ & $-11.6$ & $ 1.4$ \\
WD\,1827$-$106 & $0.14$ & $0.01$ & \hspace*{.4cm} $ 1504$ & $   29$ & $-29.2$ & $ 3.1$ & $189.1$ & $ 3.6$ & $-71.8$ & $ 6.5$ \\
WD\,1834$-$781 & $0.41$ & $0.04$ & \hspace*{.4cm} $ 1009$ & $   55$ & $-56.7$ & $ 8.8$ & $126.6$ & $ 6.9$ & $-92.5$ & $ 6.8$ \\
WD\,1840+042 & $0.07$ & $0.01$ & \hspace*{.4cm} $ 1733$ & $   20$ & $-22.5$ & $ 3.3$ & $217.2$ & $ 2.4$ & $20.2$ & $ 1.8$ \\
\hline
\end{tabular}
\end{table*}
%%%%%%%%%%%%%%%%%%%%%%%%%%%%%%%%%%%%%%%%%%%%%%%%%%%%%%%%%%
\begin{table*}[h!]
\addtocounter{table}{-1}
\caption[]
{Kinematic parameters of the $398$ SPY white dwarfs\\
continued from previous page}
\begin{tabular}{lr@{$\pm$}lr@{$\pm$}lr@{$\pm$}lr@{$\pm$}lr@{$\pm$}l}
\hline
star & \multicolumn{2}{c}{$e$} & \multicolumn{2}{c}{$J_z$} & \multicolumn{2}{c}{$U$} & \multicolumn{2}{c}{$V$} & \multicolumn{2}{c}{$W$} \\ 
 & \multicolumn{2}{c}{}& \multicolumn{2}{c}{${\rm kpc\,km\,s^{-1}}$} & \multicolumn{2}{c}{${\rm km\,s^{-1}}$} & \multicolumn{2}{c}{${\rm km\,s^{-1}}$} & \multicolumn{2}{c}{${\rm km\,s^{-1}}$} \\
\hline
WD\,2347$-$192 & $0.22$ & $0.02$ & \hspace*{.4cm} $ 2005$ & $   25$ & $53.0$ & $ 5.0$ & $251.1$ & $ 3.1$ & $ 8.3$ & $ 4.3$ \\
WD\,2348$-$244 & $0.20$ & $0.01$ & \hspace*{.4cm} $ 1661$ & $   23$ & $61.9$ & $ 3.4$ & $207.8$ & $ 2.8$ & $-23.3$ & $ 4.2$ \\
WD\,2349$-$283 & $0.05$ & $0.01$ & \hspace*{.4cm} $ 1680$ & $   18$ & $10.1$ & $ 1.0$ & $210.5$ & $ 2.2$ & $-10.6$ & $ 2.8$ \\
WD\,2350$-$083 & $0.28$ & $0.05$ & \hspace*{.4cm} $ 1716$ & $   80$ & $-84.0$ & $15.8$ & $215.2$ & $ 9.9$ & $-47.8$ & $ 5.3$ \\
WD\,2350$-$248 & $0.13$ & $0.01$ & \hspace*{.4cm} $ 1955$ & $    9$ & $21.7$ & $ 1.3$ & $244.9$ & $ 1.2$ & $-7.2$ & $ 4.9$ \\
WD\,2351$-$368 & $0.78$ & $0.06$ & \hspace*{.4cm} $  444$ & $  101$ & $76.3$ & $ 6.2$ & $55.6$ & $12.7$ & $62.8$ & $ 4.6$ \\
WD\,2354$-$151 & $0.09$ & $0.01$ & \hspace*{.4cm} $ 1810$ & $   29$ & $28.4$ & $ 3.8$ & $226.6$ & $ 3.7$ & $-7.2$ & $ 3.9$ \\
WD\,2359$-$324 & $0.66$ & $0.04$ & \hspace*{.4cm} $  696$ & $   81$ & $-125.0$ & $ 9.8$ & $87.4$ & $10.2$ & $-88.5$ & $ 4.1$ \\
WD\,2359$-$434 & $0.07$ & $0.02$ & \hspace*{.4cm} $ 1767$ & $   21$ & $-16.8$ & $ 3.0$ & $220.9$ & $ 2.7$ & $64.9$ & $ 8.7$ \\
\hline
\end{tabular}
\end{table*}

%%%%%%%%%%%%%%%%%%%%%%%%%%%%%%%%%%%%%%%%%%%%%%%%%%%%%%%%%%%%%%%%%%%%%
\end{document}